\documentclass[letterpaper, 11pt]{article}
\usepackage{graphicx}
\usepackage{amsmath}
\usepackage{amsfonts}
\usepackage{hyperref}
\usepackage{amsthm}
 \usepackage{amscd}
\usepackage{mathrsfs}
\usepackage{caption}
\usepackage{subcaption}
\usepackage{float}
\usepackage[top=1in,left=1in,right=1in,bottom=1in]{geometry}

\counterwithin{figure}{section}
\hypersetup{
	colorlinks = true,
	linkcolor  = red,
	citecolor  = blue,
	filecolor  = cyan,
	urlcolor   = magenta,}
\newcommand{\ds}{\displaystyle}

\numberwithin{equation}{section}

\title{Soliton solutions to the Sawada--Kotera equation}
\author{Tuncay Aktosun\thanks{Dedicated to the memory of Prof. Martin Klaus of Virginia Tech, a long-time friend and collaborator}\\
Department of Mathematics\\
University of Texas at Arlington\\
Arlington, TX 76019-0408, USA\\
\\
Abdon E. Choque-Rivero\\
Instituto de F\'isica y Matem\'aticas\\
Universidad Michoacana de San Nicolás de Hidalgo\\
Morelia, Michoac\'an, M\'exico\\
\\
Ivan Toledo\\
Department of Mathematics\\
University of Texas at Arlington\\
Arlington, TX 76019-0408, USA\\
\\
Mehmet Unlu\\
Department of Mathematics\\
Recep Tayyip Erdogan University\\
53100 Rize, Turkey}

\date{}

\begin{document}

\maketitle

\begin{abstract}
We consider the direct and inverse scattering problems for the third-order differential equation
in the reflectionless case.
We formulate a corresponding Riemann--Hilbert problem
using input consisting of the bound-state
poles of a transmission coefficient and
the bound-state dependency constants.
With the time-evolved dependency constants, using
the solution to the Riemann--Hilbert problem, we construct
soliton solutions to an integrable system of fifth-order nonlinear
partial differential equations. By imposing some appropriate restrictions
on the dependency constants, we show that those soliton solutions yield
soliton solutions to the Sawada--Kotera equation.

\end{abstract}



\newpage

\section{Introduction}
\label{section1}

In this paper we are interested in soliton solutions to an integrable
system of coupled fifth-order
nonlinear partial differential equations 
and, in particular, one of its special cases, namely, the
Sawada--Kotera equation  \cite{SK1974}.
We construct such solutions by solving the inverse scattering problem
for a third-order linear differential equation in the reflectionless case.
The solution to the inverse scattering problem is
obtained by 
solving a related Riemann--Hilbert problem
using input consisting of the bound-state poles of a transmission coefficient and the bound-state dependency constants.
Using the time-evolved dependency constants
in the input data set, we show that
the solution to the inverse problem yields soliton solutions to
the aforementioned integrable nonlinear system
and also soliton solutions to
the Sawada--Kotera equation.
This method explains the physical origin of the $2\mathbf N$ real parameters
appearing in the relevant
$\mathbf N$-soliton solution formula obtained by the bilinear method \cite{H1989} of Hirota,
by relating $\mathbf N$ of the real parameters to the bound-state poles and relating
the remaining $\mathbf N$ real parameters to the bound-state dependency constants.

Our paper presents the construction of soliton solutions to the Sawada--Kotera equation via the inverse scattering transform method \cite{GGKM1967}. Such
solutions to the Sawada--Kotera equation are usually obtained by using Hirota's bilinear method or a slight modification \cite{HN1997,P2001} of that method.
 Hirota's bilinear method is an effective algebraic
method to obtain soliton solutions to various integrable nonlinear partial differential equations. However, it is an ad hoc procedure, and it does not 
provide any insight or physical motivation for the construction of those soliton solutions. In particular, it does not explain how the parameters
appearing in those soliton solutions may be related to any physical quantities. On the other hand,
the method we use to construct those soliton solutions is a fundamental procedure that can be 
applied to a wide variety of other integrable
evolution equations, and it also
relates the parameters appearing in soliton solutions
to the bound-state poles and bound-state dependency constants
for the relevant ordinary linear differential operator.

In the inverse scattering transform method, a time-evolved potential is constructed from the time-evolved scattering data. A soliton solution
corresponds to a time-evolved potential corresponding to a reflectionless scattering data set. A reflectionless scattering data set comprises the bound-state information only.
The bound-state information in turn consists of 
the bound-state
poles of a transmission coefficient and a bound-state dependency constant for each bound-state pole.
For example, an $\mathbf N$-soliton solution to the Sawada--Kotera equation is a real-valued function of $x$ and $t$ and it contains
$2\mathbf N$ real parameters. As our method indicates, $\mathbf N$ of those real parameters identify the locations of the bound-state poles of a transmission coefficient
and the remaining $\mathbf N$ real parameters identify the bound-state dependency constants 
associated with the bound-state poles.

Our paper is organized as follows. In Section~\ref{section2} we introduce the fifth-order integrable system \eqref{2.6} of coupled
nonlinear partial differential equations associated with the third-order
linear equation \eqref{2.10}. We indicate how 
the Sawada--Kotera equation arises as a special case of \eqref{2.6} by uncoupling the integrable system \eqref{2.6}.
In Section~\ref{section3} we provide a summary of the direct scattering problem for \eqref{2.10}
in the reflectionless case. This is done by introducing three relevant solutions to \eqref{2.10} at each $k$-value in the complex $k$-plane. We indicate
how the left and right transmission coefficients are related to the spacial asymptotics of two of the relevant solutions to \eqref{2.10}. In Section~\ref{section4} we introduce the bound-state poles of the left transmission coefficient $T_{\text{\rm{l}}}(k),$ and we describe the bound-state dependency constant for each bound-state pole. For 
simplicity we assume that each bound-state pole is simple.
In Section~\ref{section5} we describe the solution to the inverse scattering problem for \eqref{2.10} in the reflectionless case.
This is done by providing some explicit expressions for each of the potentials $Q$ and $P$ in terms of the input data set consisting of
$T_{\text{\rm{l}}}(k),$ the bound-state poles of $T_{\text{\rm{l}}}(k),$ and the bound-state dependency constants.
In the case of the recovery of the potentials $Q$ and $P$ that depend on the parameter $t,$ the use of time-evolved dependency
constants yields soliton solutions to the corresponding integrable system. In Section~\ref{section6}, by imposing the appropriate restrictions
on a certain solution to the nonlinear system \eqref{2.6}, we obtain the $\mathbf N$-soliton solution to
the Sawada--Kotera equation \eqref{2.1}.
Finally, in Section~\ref{section7}, we illustrate the $\mathbf N$-soliton solution to the Sawada--Kotera equation when
$\mathbf N$ takes the values of $1,$ $2,$ and $3.$

\section{The integrable system and the Sawada--Kotera equation}
\label{section2}

The Sawada--Kotera equation is the fifth-order nonlinear partial differential equation given by
\begin{equation}\label{2.1}
Q_t+Q_{xxxxx}+5\,Q_x\,Q_{xx}+5\,Q\,Q_{xxx}+5\,Q^2\,Q_x=0, \qquad x,\,t\in\mathbb R,
\end{equation}
where $\mathbb R$ is the real axis and the subscripts denote the corresponding partial derivatives. It is used as a model to describe the propagation of surface water waves
in long, narrow, shallow canals, similar to the modeling used by the KdV (Korteweg--de Vries) equation \cite{KdV1895}. For both
the KdV equation and the Sawada--Kotera equation, we assume that the solutions are real valued. Hence, the quantity
$Q$ appearing in \eqref{2.1} is assumed to be real valued. Compared to the third-order nonlinear KdV
equation, the Sawada--Kotera equation takes into account higher-order nonlinear and dispersive effects.
In Section~\ref{section7} of our paper, we remark on some similar behaviors of soliton solutions to the
Sawada--Kotera equation and to the KdV equation.

The Sawada--Kotera equation is integrable in the sense of the inverse scattering transform method. This is due to the fact that it has a Lax pair, i.e. it is related to
two linear differential operators $L$ and $A$ satisfying the Lax operator equation \cite{L1968}
 \begin{equation}
 \label{2.2}
L_t+LA-AL=0,
\end{equation}
where $L_t$ is obtained by applying the partial $t$-derivative on $L.$
By using the corresponding Lax pair $(L,A)$ on the left-hand side of \eqref{2.2}, that left-hand side becomes the zero multiplication operator
when \eqref{2.1} holds. There are actually two distinct Lax pairs for the Sawada--Kotera equation \eqref{2.1}. The first pair $(L,A)$ is given by
 \begin{equation}
 \label{2.3}
 \begin{cases}
L=D^3+QD,\\
\noalign{\medskip}
A=9\,D^5+15\,Q D^3+15\,Q_x\,D^2+\big(10\,Q_{xx}+5\,Q^2\big)D,
 \end{cases}     
\end{equation}
where we have let $D:=d/dx$ and $D^n:=d^n/dx^n$ for $n\ge 2.$ The second Lax pair $(L,A)$ for \eqref{2.1} is given by 
 \begin{equation}
 \label{2.4}
 \begin{cases}
L=D^3+QD+Q_x,\\
\noalign{\medskip}
A=9\,D^5+15\,Q D^3+30\,Q_x\,D^2+\big(25\,Q_{xx}+5\,Q^2\big)D
      +\big(10\,Q_{xxx}+10\,Q \,Q_x\big).
 \end{cases}     
\end{equation}

The Lax pairs in \eqref{2.3} and \eqref{2.4} can be obtained as the two special cases of the general Lax pair $(L,A)$ given by
 \begin{equation}
 \label{2.5}
 \begin{cases}
L=D^3+QD+P,\\
\noalign{\medskip}
A=9\,D^5+15\,Q D^3+\big(15\,P+15\,Q_x\big)D^2+\big(15\,P_x+10\,Q_{xx}+5\,Q^2\big)D
      +\big(10\,P_{xx}+10\,Q \,P\big),
 \end{cases}     
\end{equation}
where the potentials $Q$ and $P$ are independent of each other
and are allowed to take complex values. The Lax pair in \eqref{2.3} is obtained from \eqref{2.5} when $Q$ is real valued and we have
$P\equiv 0,$ and the Lax pair in \eqref{2.4} is obtained when $Q$ is real valued and we have $P=Q_x$. The general Lax pair $(L,A)$ given in \eqref{2.5} satisfies the Lax
operator equation \eqref{2.2} provided the potentials $Q$ and $P$ appearing in \eqref{2.5} satisfy the
coupled system of two fifth-order nonlinear partial differential
equations
\begin{equation}
  \label{2.6}
 \begin{cases}
 Q_t+Q_{xxxxx}+5\,Q_x\,Q_{xx}+5\,Q\, Q_{xxx}+5\,Q^2\,Q_x
           +15\,Q_{xx}\,P+15\,Q_x\, P_x-30\,P \,P_x=0,\\
\noalign{\medskip}
 \begin{aligned}
P_t+P_{xxxxx}+5\,Q\, P_{xxx}+15\,Q_x \,P_{xx}+20\,Q_{xx}\,P_x+&5\,Q^2\,P_x+10\,Q_{xxx}\,P-15\,P \,P_{xx}\\
           &+10\,Q\, Q_x P-15\,(P_x)^2=0.
\end{aligned}        
\end{cases}
\end{equation}

In our paper we assume that $Q$ and $P$ belong to the Schwartz class in $x\in\mathbb R$ for each fixed $t.$
In order to uncouple the nonlinear system \eqref{2.6}, from the last three terms on the
left-hand side of the first coupled equation in \eqref{2.6} we see that
we must have
\begin{equation*}
15\,Q_{xx}\,P+15\,Q_x\, P_x-30\,P \,P_x=0,
\end{equation*}
which is equivalent to
\begin{equation}
\label{2.8}
\left[(Q_x-P)\,P\right]_x=0.
\end{equation}
Integrating both sides of \eqref{2.8} with respect to $x$ and using the fact that $Q$ and $P$ must vanish
as $x\to\pm\infty$ for each fixed $t,$ we obtain
\begin{equation*}
(Q_x-P)\,P=0,
\end{equation*}
which shows that the uncoupling in \eqref{2.6} occurs when $P\equiv 0$ or when $P=Q_x.$
For each of those two cases, one can directly verify that the left-hand side of the second coupled equation in \eqref{2.6} 
vanishes identically when the first equation holds. Hence, we confirm that the use of $P\equiv 0$ or $P=Q_x$
in \eqref{2.6} yields \eqref{2.1}.

Associated with the linear operator $L$ in \eqref{2.5}, we have the third-order ordinary linear differential equation
\begin{equation}
\label{2.10}
\psi'''+Q(x)\,\psi'+P(x)\,\psi=k^3\,\psi, \qquad x\in\mathbb R,
\end{equation}
where the prime denotes the $x$-derivative.
We suppress the dependence on $t$ for the potentials $Q$ and $P,$ and we write
$Q(x)$ and $P(x),$ respectively, instead of $Q(x,t)$ and $P(x,t).$
The adjoint equation \cite{ACTU2025} for \eqref{2.10} is given by
\begin{equation}
\label{2.11}
    \overline \psi'''+ \overline Q(x)\, \overline \psi' + \overline P(x)\, \overline \psi = k^3\,\overline \psi, \qquad x\in \mathbb R,
\end{equation}
with the adjoint potentials $\overline Q$ and $\overline P$ related to $Q$ and $P$ as
\begin{equation}
\label{2.12}
    \overline Q(x)= Q(x)^\ast, \qquad \overline P(x)= Q'(x)^\ast - P(x)^\ast, \qquad x\in \mathbb R,
\end{equation}
where we use an asterisk for complex conjugation and use an overbar to denote the quantities associated with the adjoint equation
\eqref{2.11}.
Comparing \eqref{2.10} and \eqref{2.11}, with the help of \eqref{2.12} we observe that, when the potential $Q$ is real valued and the potential
$P$ is zero, the adjoint potential $\overline Q$ is equal to $Q$ and the adjoint potential $\overline P$ becomes equal to
$Q_x.$ In that case, the equation in \eqref{2.10} reduces to
\begin{equation}
\label{2.13}
\psi'''+Q(x)\,\psi'
=k^3\,\psi, \qquad x\in\mathbb R,
\end{equation}
and the adjoint equation in \eqref{2.11} reduces to
\begin{equation}
\label{2.14}
    \overline \psi'''+Q(x)\, \overline \psi' + Q_x(x)\, \overline \psi = k^3\,\overline \psi, \qquad x\in \mathbb R.
\end{equation}
The Lax pair in \eqref{2.3} is associated with the reduced equation
\eqref{2.13} and the Lax pair in \eqref{2.4} is 
associated with the reduced adjoint equation
\eqref{2.14}. This explains why the Sawada--Kotera equation \eqref{2.1} has
two distinct Lax pairs given in \eqref{2.3} and \eqref{2.4}, respectively.

\section{The direct scattering problem in the reflectionless case}
\label{section3}

In this section we present the basic ingredients for the direct scattering problem for \eqref{2.10} in the reflectionless case. We assume that the two 
complex-valued potentials $Q$ and $P$ appearing in
\eqref{2.10} each belong to the Schwartz class in $x\in\mathbb R$ for each fixed value of $t.$
That particular direct problem
consists of the determination of the solutions to \eqref{2.10} and the left and right transmission coefficients
$T_{\text{\rm{l}}}(k)$ and $T_{\text{\rm{r}}}(k),$ respectively,
for \eqref{2.10} when the potentials $Q$ and $P$ are known.
For the description of the direct scattering problem when the reflection coefficients
are not zero, we refer the reader to \cite{ACTU2025,K1980,T2024}.

\begin{figure}[!ht]
     \centering
         \includegraphics[width=1.8in]{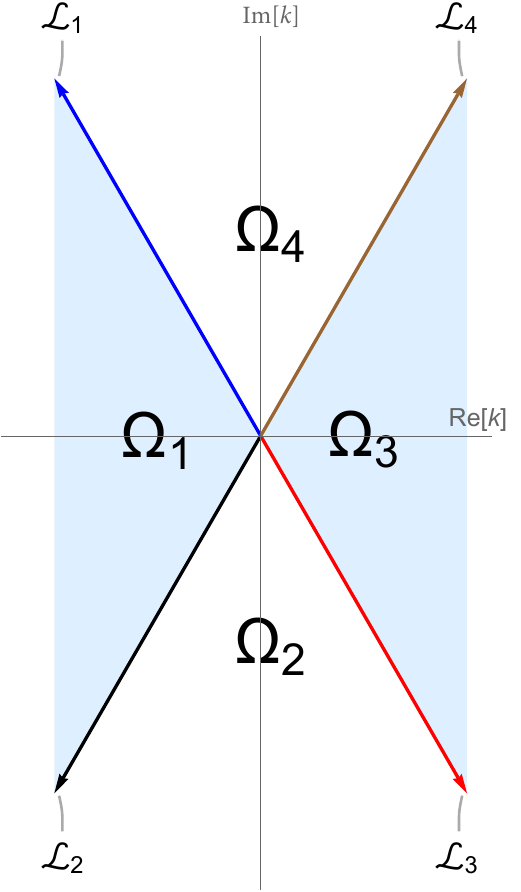}\hskip .2in
         \includegraphics[width=1.8in]{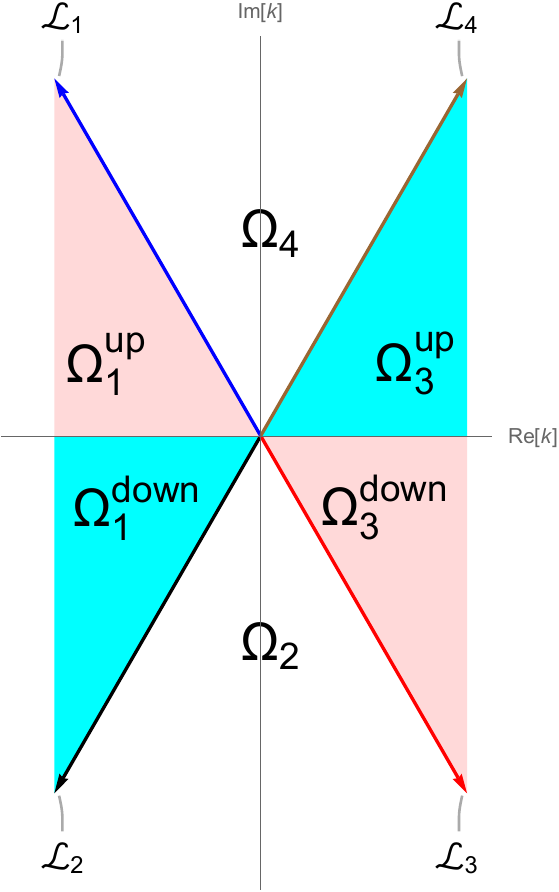}\hskip .2in
         \includegraphics[width=1.8in]{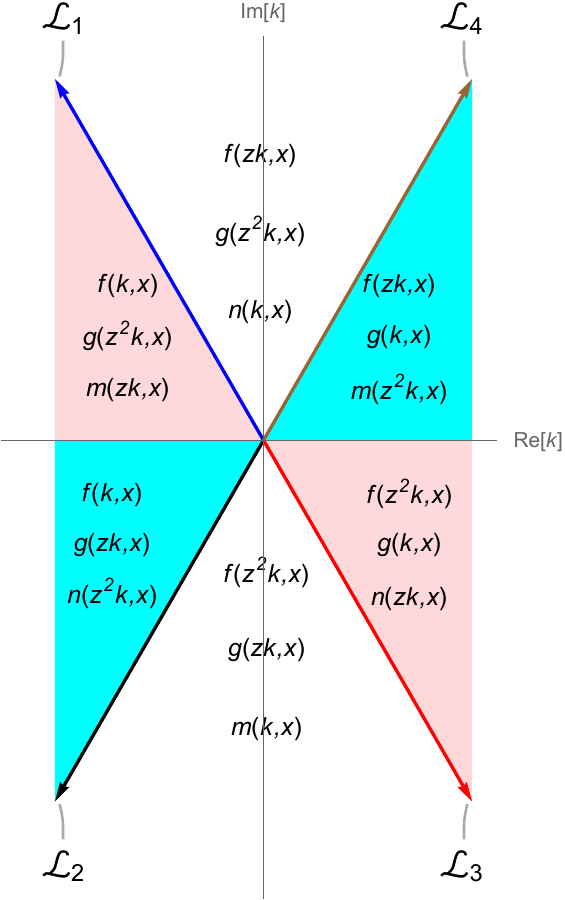}
\caption{The directed half lines
$\mathcal L_1,$ $\mathcal L_2,$ $\mathcal L_3,$ $\mathcal L_4$
and the open sectors $\Omega_1,$ $\Omega_2,$ $\Omega_3,$ $\Omega_4$ in the complex $k$-plane are shown on the left plot.
The complex $k$-plane is divided into the six sectors $\Omega^\text{\rm{up}}_1,$ $\Omega^\text{\rm{down}}_1,$ 
$\Omega_2,$ $\Omega^\text{\rm{down}}_3,$ $\Omega^\text{\rm{up}}_3,$ and $\Omega_4$ as shown on the middle plot, and the 
$k$-domains of three basic solutions to \eqref{2.10} in each of the six regions, respectively, are shown on the right plot.}
\label{figure3.1}
\end{figure}

It is convenient to divide the complex $k$-plane into four open sectors $\Omega_1,$ $\Omega_2,$ $\Omega_3,$ $\Omega_4$ as indicated
on the left plot of Figure~\ref{figure3.1} by using the directed half lines $\mathcal L_1,$ $\mathcal L_2,$ $\mathcal L_3,$ $\mathcal L_4,$ which are
parametrized as
\begin{equation}\label{3.1}
\mathcal L_1:=\{k\in\mathbb C: k=zs \text{\rm{ for }}  s\in[0,+\infty)\},
\end{equation}
\begin{equation*}
\mathcal L_2:=\{k\in\mathbb C: k=z^2s \text{\rm{ for }}  s\in[0,+\infty)\},
\end{equation*}
\begin{equation}\label{3.3}
\mathcal L_3:=\{k\in\mathbb C: k=-zs \text{\rm{ for }}  s\in[0,+\infty)\},
\end{equation}
\begin{equation*}
\mathcal L_4:=\{k\in\mathbb C: k=-z^2s \text{\rm{ for }}  s\in[0,+\infty)\}.
\end{equation*}
We use $z$ to denote the special complex number $e^{2\pi i/3},$ which is also expressed as
\begin{equation}\label{3.5}
z:=-\displaystyle\frac{1}{2}+i\,\displaystyle\frac{\sqrt{3}}{2}.
\end{equation}
The open sectors $\Omega_1,$ $\Omega_2,$ $\Omega_3,$ $\Omega_4$ are described by using
the parametrizations given by
\begin{equation*}
\Omega_1:=\left\{k\in\mathbb C: \displaystyle\frac{2\pi}{3}<\arg[k]<\displaystyle\frac{4\pi}{3}\right\},
\end{equation*}
\begin{equation*}
\Omega_2:=\left\{k\in\mathbb C: -\displaystyle\frac{2\pi}{3}<\arg[k]<-\displaystyle\frac{\pi}{3}\right\},
\end{equation*}
\begin{equation*}
\Omega_3:=\left\{k\in\mathbb C: -\displaystyle\frac{\pi}{3}<\arg[k]<\displaystyle\frac{\pi}{3}\right\},
\end{equation*}
\begin{equation*}
\Omega_4:=\left\{k\in\mathbb C: \displaystyle\frac{\pi}{3}<\arg[k]<\displaystyle\frac{2\pi}{3}\right\},
\end{equation*}
where $\arg[k]$ denotes the argument function taking values in the interval $(-2\pi/3,4\pi/3).$ We use $\overline{\Omega_1},$ $\overline{\Omega_2},$
$\overline{\Omega_3},$ $\overline{\Omega_4}$ to denote the closures of the open sectors $\Omega_1,$ $\Omega_2,$ $\Omega_3,$ $\Omega_4,$
respectively, where we recall that the closures are obtained by adding the boundaries to the corresponding open sectors.

We define the left Jost solution $f(k,x)$
and the right Jost solution $g(k,x)$ as the solutions to \eqref{2.10} with the respective spacial asymptotics given by 
\begin{equation}\label{3.10}
\begin{cases}
f(k,x)=e^{kx}\left[1+o(1)\right], \qquad x\to+\infty, \quad k\in\overline{\Omega_1},\\
\noalign{\medskip}
f'(k,x)=k\,e^{kx}\left[1+o(1)\right], \qquad x\to+\infty, \quad k\in\overline{\Omega_1},\\
\noalign{\medskip}
f''(k,x)=k^2\,e^{kx}\left[1+o(1)\right],  \qquad x\to+\infty, \quad k\in\overline{\Omega_1},
\end{cases}
\end{equation}
\begin{equation}\label{3.11}
\begin{cases}
g(k,x)=e^{kx}\left[1+o(1)\right], \qquad x\to-\infty, \quad k\in\overline{\Omega_3},\\
\noalign{\medskip}
g'(k,x)=k\,e^{kx}\left[1+o(1)\right], \qquad x\to-\infty, \quad k\in\overline{\Omega_3},\\
\noalign{\medskip}
g''(k,x)=k^2\,e^{kx}\left[1+o(1)\right],  \qquad x\to-\infty, \quad k\in\overline{\Omega_3}.
\end{cases}
\end{equation}
When the reflection coefficients for
\eqref{2.10} are all zero,
the left transmission coefficient $T_{\text{\rm{l}}}(k)$ and the right transmission coefficient $T_{\text{\rm{r}}}(k)$ are obtained by using the appropriate respective
spacial asymptotics of the Jost solutions $f(k,x)$ and $g(k,x).$ We have 
\begin{equation*}
f(k,x)=e^{kx}T_{\text{\rm{l}}}(k)^{-1}\left[1+o(1)\right], \qquad x\to-\infty, \quad k\in\overline{\Omega_1},
\end{equation*}
\begin{equation*}
g(k,x)=e^{kx}T_{\text{\rm{r}}}(k)^{-1}\left[1+o(1)\right], \qquad x\to+\infty, \quad k\in\overline{\Omega_3}.
\end{equation*}
When the potentials $Q$ and $P$ in \eqref{2.10} belong to the Schwartz class in $x\in\mathbb R,$ the transmission coefficients $T_{\text{\rm{l}}}(k)$ and $T_{\text{\rm{r}}}(k)$
have meromorphic extensions from their respective $k$-domains $\overline{\Omega_1}$ and $\overline{\Omega_3}$ to the entire complex $k$-plane.
In fact, those extensions satisfy \cite{ACTU2025}
\begin{equation}
\label{3.14}
T_{\text{\rm{r}}}(k)= \displaystyle\frac{1}{T_{\text{\rm{l}}}(k)},\qquad k\in\mathbb C.
\end{equation}

It is known \cite{ACTU2025} that \eqref{2.10} has the particular solutions $m(k,x)$ and $n(k,x)$ with the respective $k$-domains $\overline{\Omega_2}$ and
$\overline{\Omega_4},$ and in the reflectionless case they satisfy the spacial asymptotics given by 
\begin{equation*}
m(k,x)=e^{kx}\left[1+o(1)\right], \qquad  x\to-\infty, \quad k\in\overline{\Omega_2}, 
\end{equation*}
\begin{equation*}
n(k,x)=e^{kx}\left[1+o(1)\right], \qquad  x\to-\infty, \quad k\in\overline{\Omega_4}.
\end{equation*}
In the reflectionless case, we have \cite{ACTU2025} the spacial asymptotics 
\begin{equation*}
m(k,x)= e^{kx}\,T_{\text{\rm{l}}}(z^2k)^{-1}\,T_{\text{\rm{r}}}(zk)[1+o(1)],\qquad  x\to+\infty, \quad k\in\overline{\Omega_2}, 
\end{equation*}
\begin{equation*}
n(k,x)= e^{kx}\,T_{\text{\rm{l}}}(zk)^{-1}\,T_{\text{\rm{r}}}(z^2k)[1+o(1)],\qquad  x\to+\infty, \quad k\in\overline{\Omega_4}, 
\end{equation*}
where we recall that $z$ is the cube root of unity appearing in \eqref{3.5}.

Since $k$ appears as $k^3$ in \eqref{2.10}, it follows that $\psi(zk,x)$ and $\psi(z^2k,x)$ are also solutions to \eqref{2.10} whenever $\psi(k,x)$ is a solution.
The $k$-domain of $\psi(zk,x)$ is obtained from the $k$-domain of $\psi(k,x)$ by a clockwise rotation of $2\pi/3$ radians around the origin of the complex
$k$-plane. Similarly, the $k$-domain of $\psi(z^2k,x)$ is obtained from the $k$-domain of $\psi(k,x)$ by a clockwise rotation of $4\pi/3$ radians.
This allows us to determine three linearly independent solutions to \eqref{2.10} at each $k$-value in the complex $k$-plane. Toward that goal, we divide $\Omega_1$
into the open subsectors $\Omega_1^\text{\rm{up}}$ and $\Omega_1^\text{\rm{down}}$ by using the negative real axis $\mathbb R^-,$ and we divide the open sector
$\Omega_3$ into the open subsectors $\Omega_3^\text{\rm{down}}$ and $\Omega_3^\text{\rm{up}}$ by using the positive real axis $\mathbb R^+.$ On the middle plot of
Figure~\ref{figure3.1}, we show the partitioning of the complex $k$-plane into the six open sectors $\Omega_1^\text{\rm{up}},$ $\Omega_1^\text{\rm{down}},$ 
$\Omega_2,$ $\Omega_3^\text{\rm{down}},$ $\Omega_3^\text{\rm{up}},$ and $\Omega_4.$ We use $-\mathbb R^-$ and $\mathbb R^+,$ respectively, to
denote the directed half lines parametrized as
\begin{equation*}
\mathbb R^+:=\{k=s: s\in[0,+\infty)\},
\end{equation*}
\begin{equation*}
-\mathbb R^-:=\{k=-s: s\in[0,+\infty)\}.
\end{equation*}
On the right plot of Figure~\ref{figure3.1} we indicate three linearly independent solutions to \eqref{2.10} in each of the six closed
sectors $\overline{\Omega_1^\text{\rm{up}}},$ $\overline{\Omega_1^\text{\rm{down}}},$ $\overline{\Omega_2},$ $\overline{\Omega_3^\text{\rm{down}}},$
$\overline{\Omega_3^\text{\rm{up}}},$ and $\overline{\Omega_4},$ where we recall that the overbar denotes the closure of the corresponding open sector.

For fixed real values of $x$ and $t,$ the large $k$-asymptotics of the basic solutions $f(k,x),$ $g(k,x),$ $m(k,x),$ and $n(k,x)$ are, respectively, given by \cite{ACTU2025}
\begin{equation}
\label{3.21}
 f(k,x) = e^{kx}\left[1 +\displaystyle\frac{u_1(x)}{k} + \displaystyle\frac{u_2(x)}{k^2} + O\left(\displaystyle\frac{1}{k^3}\right)\right],\qquad k\to\infty  
 \text{\rm{ in }} \overline{\Omega_1},
 \end{equation}
 \begin{equation}
\label{3.22}
 g(k,x) = e^{kx}\left[1 +\displaystyle\frac{v_1(x)}{k} + \displaystyle\frac{v_2(x)}{k^2} + O\left(\displaystyle\frac{1}{k^3}\right)\right],\qquad k\to\infty 
  \text{\rm{ in }}\overline{\Omega_3}, 
 \end{equation}
 \begin{equation}\label{3.23}
m(k,x)=e^{kx}\left[1+O\left(\displaystyle\frac{1}{k}\right)\right], \qquad  k\to\infty  \text{\rm{ in }} \overline{\Omega_2},
\end{equation}
\begin{equation}\label{3.24}
n(k,x)=e^{kx}\left[1+O\left(\displaystyle\frac{1}{k}\right)\right], \qquad  k\to\infty  \text{\rm{ in }} \overline{\Omega_4},
\end{equation}
where we have defined
\begin{equation}\label{3.25}
 u_1(x):= \displaystyle\frac{1}{3}\int_x^{\infty} dy\, Q(y),\qquad x\in\mathbb R,
 \end{equation}
 \begin{equation}\label{3.26}
 u_2(x):= -\displaystyle\frac{1}{3}\int_x^{\infty} dy \left[Q'(y)-P(y)\right] + \displaystyle\frac{1}{18}\left[\int_x^{\infty} dy\, Q(y)\right]^2,
 \qquad x\in\mathbb R,
  \end{equation}
\begin{equation}\label{3.27}
 v_1(x):= -\displaystyle\frac{1}{3}\int_{-\infty}^x dy\, Q(y),\qquad x\in\mathbb R,
 \end{equation}
 \begin{equation}\label{3.28}
 v_2(x):= \displaystyle\frac{1}{3}\int_{-\infty}^x dy \left[Q'(y)-P(y)\right] + \displaystyle\frac{1}{18}\left[\int_{-\infty}^x dy\, Q(y)\right]^2, \qquad x\in\mathbb R.
  \end{equation}
Using \eqref{3.25}--\eqref{3.28}, we express the potentials $Q$ and $P$ in terms of $u_1(x)$ and $u_2(x)$ as
\begin{equation}\label{3.29}
 Q(x)= -3\,\displaystyle\frac{du_1(x)}{dx}, \qquad x\in\mathbb R,
 \end{equation}
 \begin{equation}\label{3.30}
 P(x)= 3\left[u_1(x)\,\displaystyle\frac{du_1(x)}{dx}- \displaystyle\frac{d^2u_1(x)}{dx^2}-\displaystyle\frac{du_2(x)}{dx}\right], \qquad x\in\mathbb R,
 \end{equation}
 or in terms of $v_1(x)$ and $v_2(x)$ as
 \begin{equation}\label{3.31}
 Q(x)= -3\,\displaystyle\frac{dv_1(x)}{dx}, \qquad x\in\mathbb R,
 \end{equation}
 \begin{equation}\label{3.32}
 P(x)= 3\left[v_1(x)\,\displaystyle\frac{dv_1(x)}{dx}- \displaystyle\frac{d^2v_1(x)}{dx^2}-\displaystyle\frac{dv_2(x)}{dx}\right], \qquad x\in\mathbb R.
 \end{equation}

\section{The bound states and the bound-state dependency constants}
\label{section4}

A bound state corresponds to a nontrivial solution to \eqref{2.10} which is square integrable in $x\in\mathbb R.$ If a bound state
occurs at $k=k_j$ somewhere in the complex $k$-plane, then the number of linearly independent square-integrable solutions to \eqref{2.10} at $k=k_j$
determines the multiplicity of that bound state. In our paper, we only consider simple bound states, where the multiplicity of each bound state is equal
to $1.$ We refer the reader to \cite{AE2019,AE2022,AEU2023} for the treatment of bound states with multiplicities for various differential and difference equations. 
The analysis of solutions to \eqref{2.10} at $k=0$ is more challenging. When we analyze bound states for \eqref{2.10}, 
we assume that a bound state does not occur at $k=0.$

We consider the bound states occurring at the zeros of $T_{\text{\rm{l}}}(k)^{-1}$ in $\Omega_1^\text{\rm{down}}$ or $\Omega_1^\text{\rm{up}}.$ Without loss of
generality, it is enough to consider the bound states at the poles of $T_{\text{\rm{l}}}(k)$ in $\Omega_1^\text{\rm{down}}.$ It turns out
\cite{ACTU2025} that the analysis of a
bound state at $k=k_j$ with $k_j\in\Omega_1^\text{\rm{down}}$ yields useful information about
the bound state at $k=k_j^\ast$ with 
$k_j^\ast\in\Omega_1^\text{\rm{up}}.$ We recall that we use an asterisk
to denote complex conjugation.

We define the $3$-Wronskian of three functions $F(x),$ $G(x),$ and $H(x)$ as
\begin{equation*}
\left[F(x);G(x);H(x)\right]:=\begin{vmatrix}
F(x) & G(x) & H(x)\\
F'(x) & G'(x) & H'(x)\\
F''(x) & G''(x) & H''(x)
\end{vmatrix},
\end{equation*}
where we have the determinant of the relevant $3\times 3$ matrix on the right-hand side. 
The $3$-Wronskian of any three solutions to \eqref{2.10} at any particular $k$-value is zero if and only if those three
solutions are linearly dependent. Furthermore, because of the absence of the term $\psi''$ in \eqref{2.10}, 
the $3$-Wronskian of any three solutions is independent of $x$ and its value can be evaluated at any particular
$x$-value. For example, using their asymptotics as $x\to\pm\infty,$ we evaluate the $3$-Wronskian of $f(k,x),$ $g(zk,x),$ and
$n(z^2k,x)$ in $\overline{\Omega_1^\text{\rm{down}}}$ as
\begin{equation}
\label{4.2}
\left[f(k,x);g(zk,x);n(z^2k,x)\right]=-3 z(1-z)k^3 \,T_{\text{\rm{l}}}(k)^{-1},
\qquad k\in\overline{\Omega_1^\text{\rm{down}}}.
\end{equation}

Let us assume that the nonzero complex constant $k_j$ is located in the open sector $\Omega_1^\text{\rm{down}}$ and that it
corresponds to a bound state. Then, we have
$T_{\text{\rm{l}}}(k_j)^{-1}=0.$ From \eqref{4.2} we see that the three solutions
$f(k_j,x),$ $g(zk_j,x),$ and $n(z^2k_j,x)$ to
\eqref{2.10} are linearly dependent.
This allows us \cite{ACTU2025}
to express $f(k_j,x)$ as a linear combination of $g(zk_j,x)$ and $n(z^2k_j,x)$ as
\begin{equation}
\label{4.3}
f(k_j,x)=D(k_j)\,g(zk_j,x)+W(k_j)\,n(z^2k_j,x), \qquad x\in\mathbb R,
\end{equation}
for some complex-valued constants $D(k_j)$ and $W(k_j).$

Let us divide the open sector $\Omega_1^\text{\rm{down}},$ shown on the middle plot of
Figure~\ref{figure3.1}, into
two parts, the first of which is the sector
with $\arg[k]\in (\pi,7\pi/6)$ and the second sector is described via $\arg[k]\in [7\pi/6,4\pi/3).$
The analysis in Section~3 of \cite{ACTU2025} shows that
$D(k_j)=0$
when $\arg[k_j]\in (\pi,7\pi/6),$ 
and hence 
\eqref{4.3} yields
\begin{equation}
\label{4.4}
f(k_j,x)=W(k_j) \, n(z^2k_j,x),\qquad \arg[k_j]\in \left(\pi,\displaystyle\frac{7\pi}{6}\right), \quad x\in\mathbb R,
\end{equation}
where the nonzero complex constant $W(k_j)$ corresponds to the dependency constant at the bound state with $k=k_j.$
On the other hand, again
from the analysis in Section~3 of \cite{ACTU2025} it follows that
$W(k_j)=0$
if we have $\arg[k_j]\in [7\pi/6,4\pi/3).$ In that case, from \eqref{4.3} with $W(k_j)=0$ we get
\begin{equation}
\label{4.5}
f(k_j,x)=D(k_j) \, g(zk_j,x),\qquad \arg[k_j]\in \left[\displaystyle\frac{7\pi}{6},\displaystyle\frac{4\pi}{3}\right), \quad x\in\mathbb R,
\end{equation}
with the nonzero complex constant $D(k_j)$ corresponding to the dependency constant at the bound state with $k=k_j.$
If the potentials $Q$ and $P$ in \eqref{2.10} depend also on the parameter $t,$ then the solutions 
to \eqref{2.10} also depend on $t.$ Consequently, the dependency constants
$W(k_j)$ and $D(k_j)$ appearing in \eqref{4.4} and \eqref{4.5} also depend on $t.$
For further analysis related to the bound states of \eqref{2.10}, we refer the reader to \cite{ACTU2025}.

Let us consider the case where the potentials $Q$ and $P$ depend on the time variable $t$ with the time evolution governed by the linear operator $A$
given in the second line of \eqref{2.5}. In that case, the bound-state dependency constants evolve in time as
\begin{equation}\label{4.6}
D(k_j)=E(k_j)\,e^{9(z^2-1)\,k_j^5\,t},\qquad \arg[k_j]\in \left[\displaystyle\frac{7\pi}{6},\displaystyle\frac{4\pi}{3}\right),
\end{equation}
\begin{equation*}
W(k_j)=U(k_j)\,e^{9(z-1)k_j^5t}, \qquad \arg[k_j]\in \left(\pi,\displaystyle\frac{7\pi}{6}\right),
\end{equation*}
where $E(k_j)$ and $U(k_j)$ are used to denote the values of the dependency constants $D(k_j)$ and $W(k_j),$ respectively, at $t=0.$
We refer the reader to \cite{ACTU2025} for further information on the time evolution of the bound-state dependency
constants.

\section{The inverse scattering problem in the reflectionless case}
\label{section5}

In this section we present the basic results for the inverse problem for \eqref{2.10} in the reflectionless case. In that case, the
 inverse problem 
consists of the determination of the potentials $Q$ and $P$ when the transmission coefficients $T_{\text{\rm{l}}}(k)$ and $T_{\text{\rm{r}}}(k)$ are known.
In fact, as a result of \eqref{3.14}, the right transmission coefficient $T_{\text{\rm{r}}}(k)$
is known when $T_{\text{\rm{l}}}(k)$ is specified.
In general, the potentials $Q$ and $P$ cannot be determined uniquely when we only know the transmission coefficients. For the unique 
recovery of the potentials, we need some additional information. The specification of a dependency constant or a normalization constant
at each bound state is a conventional way to recover the potentials uniquely. A bound-state dependency constant and a bound-state normalization constant are
related \cite{ACTU2025} to each other via the residue of a transmission coefficient at the bound-state pole. The use of bound-state dependency constants is convenient
when the potentials are recovered from the solution to a related Riemann-Hilbert problem. The use of bound-state normalization constants
is appropriate when the potentials are recovered from the solution to a related linear integral equation, such as a Marchenko integral equation \cite{AK2001,CS1989,DT1979,F1967,L1987,M2011}
or a
Gel'fand--Levitan integral equation \cite{CS1989,L1987,M2011}.

In the reflectionless case, we  formulate the inverse problem 
for \eqref{2.10} as a Riemann--Hilbert problem in the complex $k$-plane. Thus, the left transmission coefficient and the bound-state dependency
constants can be used as input to the Riemann-Hilbert problem, and the potentials are uniquely recovered from the solution to
the Riemann-Hilbert problem. We refer the reader to Section 4 of \cite{ACTU2025} for the details of the analysis of 
the formulation of the corresponding Riemann--Hilbert problem.
In this section, we present the relevant Riemann--Hilbert problem when a specific left transmission coefficient is used as input. The particular choice of the
left transmission coefficient is motivated by the goal of obtaining soliton solutions to the Sawada--Kotera equation \eqref{2.1}.

We choose the left transmission coefficient $T_{\text{\rm{l}}}(k)$ as
\begin{equation}
\label{5.1}
T_{\text{\rm{l}}}(k)=\displaystyle\frac{\Gamma(k)}{\Gamma(-k)},
\end{equation}
where $\Gamma(k)$ is given by
\begin{equation}
\label{5.2}
\Gamma(k):=\displaystyle\prod_{j=1}^{\mathbf N} (k+k_j)(k+k_j^\ast),\qquad k\in\mathbb C.
\end{equation}
From \eqref{5.1} and \eqref{5.2} we see that the left
transmission coefficient $T_{\text{\rm{l}}}(k)$ is defined on the whole complex $k$-plane and that it has $2{\mathbf N}$ poles located at $k=k_j$ and $k=k_j^\ast,$ respectively, for
$1\le j\le{\mathbf N}.$ 
The large $k$-asymptotics of $T_{\text{\rm{l}}}(k)$ is given by
\begin{equation*}
T_{\text{\rm{l}}}(k)=1+\displaystyle\frac{2\,\Sigma_{\mathbf N}}{k}+\displaystyle\frac{2\,\Sigma_{\mathbf N}^2}{k^2}+O\left(\displaystyle\frac{1}{k^3}\right),\qquad k\to\infty  \text{\rm{ in }}\mathbb C, 
 \end{equation*}
where we have defined
\begin{equation}\label{5.4}
\Sigma_{\mathbf N}:=\sum_{j=1}^{\mathbf N} \left(k_j+k_j^\ast\right).
 \end{equation}

The quantity $Q$ appearing in the Sawada--Kotera equation \eqref{2.1} is real valued. 
Since we would like to relate \eqref{5.1} eventually to
\eqref{2.1}, we 
are interested in the consideration of
the potentials $Q$ and $P$ appearing in \eqref{2.10} only when they are real valued.
Since the original $k$-domain of $T_{\text{\rm{l}}}(k)$ is
the sector
$\overline{\Omega_1},$ the restriction of $Q$ to the real values 
requires \cite{ACTU2025} that each of the $k_j$-values in \eqref{5.2} must occur on the directed half line
with $\arg[k]=7\pi/6.$ Thus, we have a restriction on the locations of the
bound-state poles. Since we only consider simple bound states, the restriction can be presented as
\begin{equation}
\label{5.5}
k_j=iz\eta_j, \qquad 1\le j\le {\mathbf N},
\end{equation}
where we recall that $z$ is the special complex constant in \eqref{3.5}.
Without loss of generality, we assume that $0<\eta_1<\dots<\eta_{\mathbf N}.$

\begin{figure}[!ht]
     \centering
         \includegraphics[width=2.in,height=3.4in]{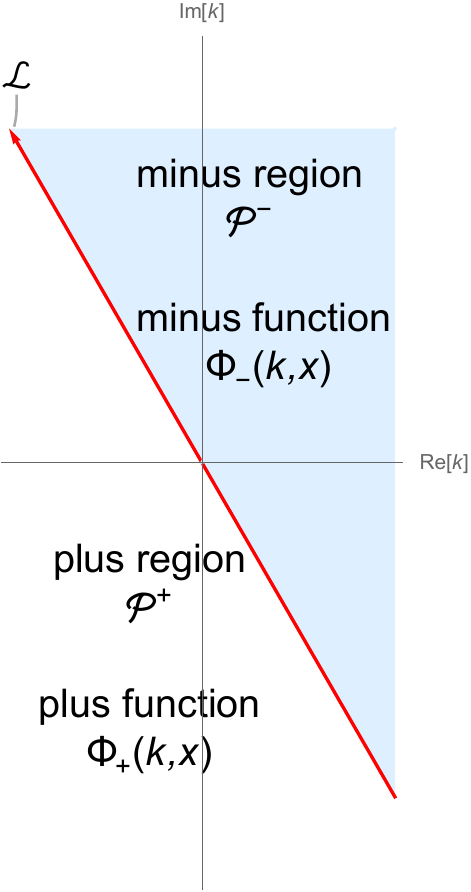} 
\caption{The plus and minus regions in the complex $k$-plane are separated by the directed full line $\mathcal L,$ and the
 plus and minus functions have their respective $k$-domains as indicated.}
\label{figure5.1}
\end{figure}

We refer the reader to Section 4 of \cite{ACTU2025} for the formulation of the Riemann--Hilbert problem 
in the presence of reflection coefficients.
Here, we
present the basic steps of the formulation of the Riemann-Hilbert problem  in the reflectionless case and
we also describe  the basic steps to
solve that Riemann--Hilbert problem.
For the details of the steps outlined here, we also refer the reader
to Section 4 of \cite{ACTU2025}.
Using the basic solutions $f(k,x),$ $g(k,x),$ $m(k,x),$ and $n(k,x)$ introduced in Section~\ref{section3},
we form the
solutions $\Phi_+(k,x)$ and $\Phi_-(k,x)$ to
\eqref{2.10} as
\begin{equation}
\label{5.6}
    \Phi_+(k,x) := 
    \begin{cases}
        T_\text{\rm{l}}(k)\, f(k,x), \qquad k\in \overline{\Omega_1},\\
                \noalign{\medskip}
        m(k,x), \qquad k\in \overline{\Omega_2},
    \end{cases}
\end{equation}
\begin{equation}
\label{5.7}
    \Phi_-(k,x) := 
    \begin{cases}
        g(k,x), \qquad k\in \overline{\Omega_3},\\
        \noalign{\medskip}
        n(k,x), \qquad k\in \overline{\Omega_4}.
    \end{cases}
\end{equation}
Using the directed half lines $\mathcal L_1$ and $\mathcal L_3$ defined in \eqref{3.1} and \eqref{3.3}, respectively, we obtain the directed full line
$\mathcal L$ via $\mathcal L:=\mathcal L_1\cup (-\mathcal L_3),$ where we recall that
$-\mathcal L_3$ is obtained from $\mathcal L_3$ by changing its direction.
The parametrization of $\mathcal L$ is given by
\begin{equation*}
\mathcal L:=\{k\in\mathbb C: k=zs \text{\rm{ for }} s\in(-\infty,+\infty)\}.
\end{equation*}
The directed line $\mathcal L$ divides the complex $k$-plane into two half planes $\mathcal P^+$ and $\mathcal P^-$ as shown in Figure~\ref{figure5.1}.
The open left-half complex plane $\mathcal P^+$ and the open right-half complex plane $\mathcal P^-$ can be parameterized as
\begin{equation*}
\mathcal P^+:=\{k\in\mathbb C: k=zs \text{\rm{ for }} s\in\mathbb C^+\},
\end{equation*}
\begin{equation*}
\mathcal P^-:=\{k\in\mathbb C: k=zs \text{\rm{ for }} s\in\mathbb C^-\},
\end{equation*}
where we use $\mathbb C^+$ and $\mathbb C^-$ to denote the upper-half and lower-half complex planes.
We refer to $\mathcal P^+$ as the plus region and refer to $\mathcal P^-$ as the minus region, as indicated in Figure~\ref{figure5.1}. We use
$\overline{\mathcal P^+}$ and $\overline{\mathcal P^-}$ to denote their closures. Hence, we have $\overline{\mathcal P^+}:=\mathcal P^+\cup\mathcal L$
and $\overline{\mathcal P^-}:=\mathcal P^-\cup\mathcal L.$

We remark that the quantities $\Phi_+(k,x)$ and $\Phi_-(k,x)$ in \eqref{5.6} and \eqref{5.7}, respectively, are
certain specific solutions to \eqref{2.10}. We proceed to set up our Riemann--Hilbert
problem. It is known \cite{ACTU2025} in the reflectionless case that $\Phi_+(k,x)$ and $\Phi_-(k,x)$ satisfy
\begin{equation}\label{5.11}
\Phi_+(k,x)=\Phi_-(k,x), \qquad k\in\mathcal L.
\end{equation}
When \eqref{5.1}, \eqref{5.2}, and \eqref{5.5} hold, it is known \cite{ACTU2025} that, for each fixed $x\in\mathbb R,$ the quantity $\Phi_+(k,x)$ is meromorphic in
$k\in\mathcal P^+$ with simple poles located at $k=k_j$ and $k=k_j^\ast$ 
for $1\le j\le\mathbf N.$ On the other hand, for each fixed $x\in\mathbb R,$ the quantity $\Phi_-(k,x)$ is
analytic in $k\in\mathcal P^-.$ From \eqref{5.11} we conclude that $\Phi_+(k,x)$ is the meromorphic continuation of $\Phi_-(k,x)$ from $k\in\mathcal P^-$ to
$k\in\mathcal P^+,$ and that $\Phi_-(k,x)$ is the analytic continuation of $\Phi_+(k,x)$ from $k\in\mathcal P^+$ to $k\in\mathcal P^-.$ Multiplying both sides
of \eqref{5.11} by $e^{-kx}\,\Gamma(-k),$ we get
\begin{equation}\label{5.12}
e^{-kx}\,\Gamma(-k)\,\Phi_+(k,x)=e^{-kx}\,\Gamma(-k)\,\Phi_-(k,x), \qquad k\in \mathcal L.
\end{equation}
For each fixed $x\in\mathbb R,$ the left-hand side of \eqref{5.12} is analytic in $k\in\mathcal P^+$ and the right-hand side is analytic in
$k\in\mathcal P^-.$ Hence, those two sides are analytic continuations of each other. Consequently, each side of \eqref{5.12} is entire
with their respective analytic continuations in $k\in\mathbb C.$ With the help of
\eqref{3.21}--\eqref{3.29}, by using the generalized Liouville theorem \cite{R1987}, we conclude that each side of \eqref{5.12} is equal to a monic
polynomial in $k$ of degree $2\mathbf N,$ where the coefficients may depend on $x$ and $t.$ By suppressing the $t$-dependence in our notation, we obtain the solution to our Riemann--Hilbert problem posed in \eqref{5.11} as
\begin{equation}\label{5.13}
\Phi_+(k,x)=e^{kx}\,\displaystyle\frac{k^{2\mathbf N}+V(k)\,\mathbf A(x)}{\Gamma(-k)}, \qquad k\in \overline{\mathcal P^+}, \quad x\in\mathbb R,
\end{equation}
\begin{equation}\label{5.14}
\Phi_-(k,x)=e^{kx}\,\displaystyle\frac{k^{2\mathbf N}+V(k)\,\mathbf A(x)}{\Gamma(-k)}, \qquad k\in \overline{\mathcal P^-}, \quad x\in\mathbb R,
\end{equation}
where $V(k)$ is the row vector with $2\mathbf N$ components defined as
\begin{equation*}
V(k):=\begin{bmatrix}
k^{2\mathbf N-1} &k^{2\mathbf N-2} & \cdots & k &1
\end{bmatrix},\qquad k\in\mathbb C,
\end{equation*}
and $\mathbf A(x)$ is a column vector with $2\mathbf N$ entries that are functions of $x$ and $t.$
Because of the aforementioned meromorphic extensions, each of \eqref{5.13} and \eqref{5.14} holds when $k\in\mathbb C.$
We write $\mathbf A(x)$ in terms of its components as
\begin{equation*}
\mathbf A(x)=\begin{bmatrix}
A_{2\mathbf N-1}(x)\\
A_{2\mathbf N-2}(x)\\
\vdots\\
A_1(x)\\
A_0(x)
\end{bmatrix},\qquad x\in\mathbb R.
\end{equation*}
We emphasize that we suppress the $t$-dependence in our notation for $\mathbf A(x)$ and its components $A_j(x)$ for $0\le j\le 2\mathbf N-1.$

We remark that the solutions $\Phi_+(k,x)$ and $\Phi_-(k,x)$ appearing in \eqref{5.13} and \eqref{5.14} comprise the general solution to the
Riemann--Hilbert problem \eqref{5.11}. In other words, those solutions contain $2\mathbf N$ arbitrary constants not depending on $k$ but depending on $x$ and
$t$ through the scalar functions $A_j(x)$ for $0\le j\le 2\mathbf N-1.$  From \eqref{5.13}, with the help of \eqref{5.1} and the first line of \eqref{5.6}, we get
\begin{equation}\label{5.17}
f(k,x)=e^{kx}\,\displaystyle\frac{k^{2\mathbf N}+V(k)\,\mathbf A(x)}{\Gamma(k)}, \qquad k\in\overline{\Omega_1}, \quad x\in \mathbb R.
\end{equation}
From \eqref{5.13} and the second line of \eqref{5.6}, we obtain
\begin{equation}\label{5.18}
m(k,x)=e^{kx}\,\displaystyle\frac{k^{2\mathbf N}+V(k)\,\mathbf A(x)}{\Gamma(-k)}, \qquad k\in\overline{\Omega_2}, \quad x\in \mathbb R.
\end{equation}
Similarly, from \eqref{5.14}, with the help of \eqref{5.7}, we have 
\begin{equation}\label{5.19}
g(k,x)=e^{kx}\,\displaystyle\frac{k^{2\mathbf N}+V(k)\,\mathbf A(x)}{\Gamma(-k)}, \qquad k\in\overline{\Omega_3}, \quad x\in \mathbb R,
\end{equation}
\begin{equation}\label{5.20}
n(k,x)=e^{kx}\,\displaystyle\frac{k^{2\mathbf N}+V(k)\,\mathbf A(x)}{\Gamma(-k)}, \qquad k\in\overline{\Omega_4}, \quad x\in \mathbb R.
\end{equation}
With the help of \eqref{5.2}, we observe that 
the right-hand sides in \eqref{5.18}--\eqref{5.20} are analytic in their respective $k$-domains
and that the right-hand side in \eqref{5.17} is meromorphic in its $k$-domain.
Thus, we have confirmed the aforementioned
meromorphic and analytic properties stated below \eqref{5.11}.

In order to get a unique solution to \eqref{5.11}, we use the dependency constants $D(k_j)$ and $D(k_j^\ast)$ at the bound states at $k=k_j$ and 
$k=k_j^\ast,$ respectively, as follows. We recall that $D(k_j)$ appears in \eqref{4.5} and \eqref{4.6}.
It is known \cite{ACTU2025} that when $k_j$ has the form given in \eqref{5.5}, the bound-state dependency
constants $D(k_j)$ and $D(k_j^\ast)$ can be written in simpler forms by introducing the modified bound-state dependency constant $\gamma_j$
by letting
\begin{equation}\label{5.21}
\gamma_j:=-\displaystyle\frac{\Gamma(k_j)\,E(k_j)}{\Gamma(-zk_j)}, \qquad 1\le j\le \mathbf N, 
\end{equation}
where we recall that $E(k_j)$ and $\Gamma(k)$ are the quantities appearing in \eqref{4.6} and \eqref{5.2}, respectively.
To use the bound-state dependency constants $D(k_j)$ and $D(k_j^\ast)$ in the solution to the Riemann--Hilbert problem \eqref{5.11}, we proceed as
follows. With the help of \eqref{4.6}, \eqref{5.17}, and \eqref{5.19}, from \eqref{4.5} we obtain
\begin{equation}\label{5.22}
e^{k_jx}\, \displaystyle\frac{k_j^{2\mathbf N}+V(k_j)\,\mathbf A(x)}{\Gamma(k_j)}
=E(k_j)\,\,e^{9(z^2-1)k_j^5t} e^{zk_jx}\,\displaystyle\frac{(zk_j)^{2\mathbf N}+V(zk_j)\,\mathbf A(x)}{\Gamma(-zk_j)}, \qquad 1\le j\le \mathbf N.
\end{equation}
The analog of \eqref{4.5} at $k=k_j^\ast$ is given by \cite{ACTU2025}
\begin{equation}\label{5.23}
f(k_j^\ast,x)=D(k_j^\ast)\, g(z^2k_j^\ast,x), \qquad \arg[k_j^\ast]\in\left(\displaystyle\frac{2\pi}{3},\displaystyle\frac{5\pi}{6}\right],
\end{equation}
and the analog of \eqref{4.6} at $k=k_j^\ast$ is given by \cite{ACTU2025}
\begin{equation}\label{5.24}
D(k_j^\ast)=E(k_j^\ast)\,e^{9(z-1)\,(k_j^\ast)^5\,t}, \qquad \arg[k_j^\ast]\in\left(\displaystyle\frac{2\pi}{3},\displaystyle\frac{5\pi}{6}\right].
\end{equation} 
With the help of \eqref{5.17}, \eqref{5.19}, and \eqref{5.24}, from \eqref{5.23} we get
\begin{equation}\label{5.25}
\begin{split}
e^{k_j^\ast x}\,&\displaystyle\frac{(k_j^\ast)^{2\mathbf N}+V(k_j^\ast)\,\mathbf A(x)}{\Gamma(k_j^\ast)}\\
&\phantom{xxxxx}=E(k_j^\ast)\,e^{9(z-1)(k_j^\ast)^5t}\,e^{z^2k_j^\ast x}\,
\displaystyle\frac{(z^2k_j^\ast)^{2\mathbf N}+V(z^2k_j^\ast)
\,\mathbf A(x)}{\Gamma(-z^2k_j^\ast)}, \qquad 1\le j\le \mathbf N.
\end{split}
\end{equation}
Let us introduce the quantity $\chi_j,$ which is an exponential function of $x$ and $t,$ as
\begin{equation}\label{5.26}
\chi_j:=e^{(z-1)k_j x+9(z^2-1)k_j^5 t}, \qquad 1\le j\le \mathbf N.
\end{equation}
When $k_j$ has the form given in \eqref{5.5}, from \eqref{5.26} we conclude that 
\begin{equation*}
\chi_j^\ast=\chi_j, \qquad 1\le j\le \mathbf N,
\end{equation*}
which indicates that each $\chi_j$ is real valued.
Similarly, when $k_j$ is as in \eqref{5.5}, with the help of \eqref{4.6} and \eqref{5.2}
where the quantities $\Gamma(k)$ and $E(k_j)$ appear, respectively, we get
\begin{equation}\label{5.28}
\displaystyle\frac{\Gamma(k_j^\ast)\,E(k_j^\ast)}{\Gamma(-z^2k_j^\ast)}=\left(\displaystyle\frac{\Gamma(k_j)\,E(k_j)}{\Gamma(-zk_j)}\right)^\ast, \qquad 1\le j\le \mathbf N.
\end{equation}
Then, by comparing \eqref{5.28} with the 
right-hand side of \eqref{5.21}, we 
conclude that the modified bound-state dependency constant at $k=k_j^\ast$ is equal to
$\gamma_j^\ast,$ which is the complex conjugate of the modified bound-state dependency constant
$\gamma_j$ at $k=k_j$ appearing in \eqref{5.21}.
Consequently, we can rewrite \eqref{5.22} and \eqref{5.25}, respectively, as
\begin{equation}\label{5.29}
k_j^{2\mathbf N}+V(k_j)\,\mathbf A(x)=-\gamma_j\,\chi_j\left[(zk_j)^{2\mathbf N}+V(zk_j)\,\mathbf A(x)\right], \qquad 1\le j\le \mathbf N,
\end{equation}
\begin{equation}\label{5.30}
(k_j^\ast)^{2\mathbf N}+V(k_j^\ast)\,\mathbf A(x)=-\gamma_j^\ast\,\chi_j\left[(z^2k_j^\ast)^{2\mathbf N}
+V(z^2k_j^\ast)\,\mathbf A(x)\right], \qquad 1\le j\le \mathbf N.
\end{equation}

We remark that \eqref{5.29} and \eqref{5.30} comprise a linear algebraic system of $2\mathbf N$
equations for the $2\mathbf N$ unknowns $A_j(x)$ with $0\le j \le 2\mathbf N-1.$ 
We write that linear algebraic system in the matrix form as
\begin{equation}\label{5.31}
\mathbf M(x)\, \mathbf A(x)=-\mathbf B(x),
\end{equation}
where we have defined the $2\mathbf N\times 2\mathbf N$ matrix $\mathbf M(x)$ and the column vector $\mathbf B(x)$ with $2\mathbf N$ entries as
\begin{equation}\label{5.32}
\mathbf M(x):=\begin{bmatrix}
m_{2\mathbf N-1}(k_1) & m_{2\mathbf N-2}(k_1) & \cdots & m_1(k_1) & m_0(k_1)\\
m_{2\mathbf N-1}(k_1^\ast) & m_{2\mathbf N-2}(k_1^\ast) & \cdots & m_1(k_1^\ast) & m_0(k_1^\ast)\\
\vdots & \vdots & \ddots & \vdots & \vdots \\
m_{2\mathbf N-1}(k_{\mathbf N}) & m_{2\mathbf N-2}(k_{\mathbf N}) & \cdots & m_1(k_{\mathbf N}) & m_0(k_{\mathbf N})\\
m_{2\mathbf N-1}(k_{\mathbf N}^\ast) & m_{2\mathbf N-2}(k_{\mathbf N}^\ast) & \cdots & m_1(k_{\mathbf N}^\ast) & m_0(k_{\mathbf N}^\ast)
\end{bmatrix},
\end{equation}
\begin{equation}\label{5.33}
\mathbf B(x):=\begin{bmatrix}
m_{2\mathbf N}(k_1)\\
m_{2\mathbf N}(k_1^\ast)\\
\vdots\\
m_{2\mathbf N}(k_{\mathbf N})\\
m_{2\mathbf N}(k_{\mathbf N}^\ast)
\end{bmatrix}.
\end{equation}
The entries appearing in \eqref{5.32} and \eqref{5.33} are given as
\begin{equation}\label{5.34}
\begin{cases}
m_l(k_j):=k_j^l+(zk_j)^l\,\gamma_j\,\chi_j,
\qquad 1\le j\le \mathbf N, \quad 0\le l\le 2\mathbf N,
\\
\noalign{\medskip}
m_l(k_j^\ast):=(k_j^\ast)^l+(z^2k_j^\ast)^l\,\gamma_j^\ast\,\chi_j,\qquad 1\le j\le \mathbf N, \quad 0\le l\le 2\mathbf N.
\end{cases}
\end{equation}
From \eqref{5.34} we see that we have
\begin{equation*}
m_l(k_j^\ast)=m_l(k_j)^\ast,
\qquad 1\le j\le \mathbf N, \quad 0\le l\le 2\mathbf N.
\end{equation*}
We emphasize that we suppress the $t$-dependence in our notation for $\mathbf M(x),$ $\mathbf A(x),$ and $m_l(k_j).$
It turns out that the potential $Q$ is determined
by $A_{2\mathbf N-1}(x)$ alone and the potential $P$ is determined by $A_{2\mathbf N-1}(x)$ and $A_{2\mathbf N-2}(x)$ only. The remaining entries 
$A_j(x)$ for $0\le j \le 2\mathbf N-3,$
along with
$A_{2\mathbf N-1}(x)$ and $A_{2\mathbf N-2}(x),$ 
 are used to construct the basic solutions $f(k,x),$ $g(k,x),$ $m(k,x),$ and $n(k,x)$ to \eqref{2.10} given in \eqref{5.17}--\eqref{5.20}.
It is understood that when $\mathbf N=1,$ we only have $A_1(x)$ and $A_0(x)$ and those two quantities determine
both the potentials and the basic solutions. 

In order to obtain $A_j(x)$ for $0\le j \le 2\mathbf N-1,$ we solve the linear system in \eqref{5.31} and obtain $\mathbf A(x)$ as
\begin{equation}\label{5.36}
\mathbf A(x)=-\mathbf M(x)^{-1}\mathbf B(x).
\end{equation}
From \eqref{5.36} we recover $A_{2\mathbf N-1}(x)$ and $A_{2\mathbf N-2}(x),$ respectively, as
\begin{equation}\label{5.37}
A_{2\mathbf N-1}(x)=-\begin{bmatrix}
1&0&0& \cdots &0 
\end{bmatrix}\mathbf M(x)^{-1}\mathbf B(x),
\end{equation}
\begin{equation}\label{5.38}
A_{2\mathbf N-2}(x)=-\begin{bmatrix}
0&1&0& \cdots &0 
\end{bmatrix}\mathbf M(x)^{-1}\mathbf B(x),
\end{equation}
where the row vector in \eqref{5.37} has $1$ in its first entry and the zeros in each of the remaining $2\mathbf N-1$ entries. Similarly, the row vector in 
\eqref{5.38} has $1$ in its second entry and the zeros in all the remaining entries. By using (15) on p.12 of \cite{CH1989}, from \eqref{5.37} and
\eqref{5.38} we obtain $A_{2\mathbf N-1}(x)$ and $A_{2\mathbf N-2}(x)$ as the ratio of two determinants as
\begin{equation*}
A_{2\mathbf N-1}(x)=\displaystyle\frac{\det\begin{bmatrix}\begin{array}{c | c c c c c} 
	0 & 1&0&0& \cdots&0\\ 
	\hline 
	\mathbf B(x) & && &\mathbf M(x)&
\end{array}
\end{bmatrix}}{\det\begin{bmatrix}\mathbf M(x)\end{bmatrix}},
\end{equation*}
\begin{equation*}
A_{2\mathbf N-2}(x)=\displaystyle\frac{\det\begin{bmatrix}\begin{array}{c | c c c c c} 
	0 & 0&1&0& \cdots&0\\ 
	\hline 
	\mathbf B(x) &&&&\mathbf M(x) &
\end{array}
\end{bmatrix}}{\det\begin{bmatrix}\mathbf M(x)\end{bmatrix}}.
\end{equation*}
Alternatively, by using Cramer's rule, from \eqref{5.31} we get
\begin{equation}\label{5.41}
A_{2\mathbf N-1}(x)=-\displaystyle\frac{\det[\mathbf M_1(x)]}{\det[\mathbf M(x)]},
\end{equation}
\begin{equation}\label{5.42}
A_{2\mathbf N-2}(x)=-\displaystyle\frac{\det[\mathbf M_2(x)]}{\det[\mathbf M(x)]}.
\end{equation}
Here, $\mathbf M_1(x)$ denotes the $2\mathbf N\times 2\mathbf N$ matrix obtained by replacing the first column of the matrix $\mathbf M(x)$ with the
column vector $\mathbf B(x),$ and $\mathbf M_2(x)$ denotes the $2\mathbf N\times 2\mathbf N$ matrix obtained by replacing the second column of
$\mathbf M(x)$ with $\mathbf B(x).$

In order to solve the inverse scattering problem for \eqref{2.10} in the reflectionless case, we recall that we use the input data set
$\{\eta_j,\gamma_j\}_{j=1}^{\mathbf N},$ which allows a 
unique determination of the potentials $Q$ and $P.$ From \eqref{5.34}, we see that all entries in the matrix
$\mathbf M(x)$ and all components in the column vector $\mathbf B(x)$ are uniquely determined by the elements of the set
$\{\eta_j,\gamma_j\}_{j=1}^{\mathbf N}.$ Hence, as shown in \eqref{5.36} we uniquely determine $\mathbf A(x)$ by using the input data set
$\{\eta_j,\gamma_j\}_{j=1}^{\mathbf N}.$ In particular, we determine $A_{2\mathbf N-1}(x)$ and $A_{2\mathbf N-2}(x)$ uniquely in terms of the
elements in the input set $\{\eta_j,\gamma_j\}_{j=1}^{\mathbf N}.$

Having determined $A_{2\mathbf N-1}(x)$ and $A_{2\mathbf N-2}(x)$ uniquely by using the input data set $\{\eta_j,\gamma_j\}_{j=1}^{\mathbf N},$ we proceed
to determine the potentials $Q$ and $P$ appearing in \eqref{2.10}. Using \eqref{5.2}, \eqref{5.17}, and \eqref{5.19}, we obtain 
the asymptotics of $f(k,x)$ as $k\to \infty \text{\rm{ in }} \overline{\Omega_1}$ as
\begin{equation}\label{5.43}
f(k,x)
=e^{kx}\left[1+\displaystyle\frac{A_{2\mathbf N-1}(x)-\Sigma_{\mathbf N}}{k}
   +\displaystyle\frac{A_{2\mathbf N-2}(x)-\Sigma_{\mathbf N} A_{2\mathbf N-1}(x)+\Pi_{\mathbf N}}{k^2}+O\left(\displaystyle\frac{1}{k^3}\right)\right],
\end{equation}
and 
the asymptotics of $g(k,x)$ as $k\to \infty \text{\rm{ in }} \overline{\Omega_3}$ as
\begin{equation}\label{5.44}
g(k,x)=e^{kx}\left[1+\displaystyle\frac{A_{2\mathbf N-1}(x)+\Sigma_{\mathbf N}}{k}
 +\displaystyle\frac{A_{2\mathbf N-2}(x)+\Sigma_{\mathbf N} A_{2\mathbf N-1}(x)+\Pi_{\mathbf N}}{k^2}+O\left(\displaystyle\frac{1}{k^3}\right)\right].
\end{equation}
We note that $\Pi_{\mathbf N}$ appearing in \eqref{5.43} and \eqref{5.44} is the constant defined as
\begin{equation*}
\begin{split}
\Pi_{\mathbf N}:=&k_1\left(k_1+k_1^\ast+\cdots+k_{\mathbf N}+k_{\mathbf N}^\ast \right)+k_1^\ast\left(k_1^\ast+k_2+\cdots+k_{\mathbf N}+k_{\mathbf N}^\ast \right)\\
       &+k_2\left(k_2+k_2^\ast+\cdots+k_{\mathbf N}+k_{\mathbf N}^\ast \right)+\cdots
       +k_{\mathbf N}\left(k_{\mathbf N}+k_{\mathbf N}^\ast\right)+ k_{\mathbf N}^\ast\left(k_{\mathbf N}^\ast\right),
\end{split}
\end{equation*}
and we recall that $\Sigma_{\mathbf N}$ is the constant defined in \eqref{5.4}. Comparing \eqref{3.21} with \eqref{5.43} and comparing \eqref{3.22} with
\eqref{5.44}, we obtain
\begin{equation}\label{5.46}
u_1(x)=A_{2\mathbf N-1}(x)-\Sigma_{\mathbf N},
\end{equation}
\begin{equation}\label{5.47}
v_1(x)=A_{2\mathbf N-1}(x)+\Sigma_{\mathbf N},
\end{equation}
\begin{equation}\label{5.48}
u_2(x)=A_{2\mathbf N-2}(x)-\Sigma_{\mathbf N} A_{2\mathbf N-1}(x)+\Pi_{\mathbf N},
\end{equation}
\begin{equation}\label{5.49}
v_2(x)=A_{2\mathbf N-2}(x)+\Sigma_{\mathbf N} A_{2\mathbf N-1}(x)+\Pi_{\mathbf N}.
\end{equation}
With the help of \eqref{3.29}--\eqref{3.32} and \eqref{5.46}--\eqref{5.49}, we get
\begin{equation}\label{5.50}
Q(x)=-3\,\displaystyle\frac{dA_{2\mathbf N-1}(x)}{dx},
\end{equation}
\begin{equation}\label{5.51}
P(x)=3\left(A_{2\mathbf N-1}(x)\,\displaystyle\frac{dA_{2\mathbf N-1}(x)}{dx}-\displaystyle\frac{d^2A_{2\mathbf N-1}(x)}{dx^2}-\displaystyle\frac{dA_{2\mathbf N-2}(x)}{dx}\right).
\end{equation}
Since $A_{2\mathbf N-1}(x)$ and $A_{2\mathbf N-2}(x)$ are uniquely determined by the input data set $\{\eta_j,\gamma_j\}_{j=1}^\mathbf N,$ we see from
\eqref{5.50} and \eqref{5.51} that the same input data set uniquely determines the potentials $Q$ and $P$ appearing in \eqref{2.10}.
With the help of the first lines of \eqref{3.10} and \eqref{3.11}, from \eqref{5.43} and \eqref{5.44} we
obtain the spacial asymptotics given by 
\begin{equation}\label{5.52}
\displaystyle\lim_{x\to\pm\infty} A_{2\mathbf N-1}(x)=\pm \Sigma_{\mathbf N},
\end{equation}
\begin{equation}\label{5.53}
\displaystyle\lim_{x\to\pm\infty} A_{2\mathbf N-2}(x)=\Sigma_{\mathbf N}^2-\Pi_{\mathbf N}.
\end{equation}

We remark that the determinants of the
matrices $\mathbf M(x),$ $\mathbf M_1(x),$ 
and $\mathbf M_2(x)$
appearing in \eqref{5.41} and \eqref{5.42}
have various useful properties. Next, we elaborate on some of those properties.
For example, the determinant of
$\mathbf M_1(x)$ is determined
in terms of the determinant of
$\mathbf M(x)$ as
\begin{equation}\label{5.54}
\det[\mathbf M_1(x)]=\Sigma_{\mathbf N}\det[\mathbf M(x)]+\displaystyle\frac{d\det[\mathbf M(x)]}{dx},
\end{equation}
which enables us to write \eqref{5.41} in the equivalent form as
\begin{equation}\label{5.55}
A_{2\mathbf N-1}(x)=-
\Sigma_{\mathbf N}-\displaystyle\frac{1}{\det[\mathbf M(x)]}\displaystyle\frac{d\det[\mathbf M(x)]}{dx}.
\end{equation}
Consequently, using \eqref{5.55} on the right-hand side of \eqref{5.50}, we express
$Q$ in terms of 
 the determinant of
$\mathbf M(x)$ as
\begin{equation}\label{5.56}
Q(x)=3\,\displaystyle\frac{d}{dx}\left(\displaystyle\frac{1}
{\det[\mathbf M(x)]}\,
\displaystyle\frac{d\det[\mathbf M(x)]}{dx}\right).
\end{equation}

As seen from \eqref{5.29} and \eqref{5.32}--\eqref{5.34}, the appearance of 
the parameters 
$x$ and $t$ is restricted through the quantities
$\chi_j$ appearing in \eqref{5.26}.
In fact, from \eqref{5.32}--\eqref{5.34} we observe that each of
the determinants of the
matrices $\mathbf M(x),$ $\mathbf M_1(x),$ 
and $\mathbf M_2(x)$
is a polynomial of degree $2\mathbf N$ in the
$\mathbf N$ real-valued variables
$\chi_1,\dots,\chi_{\mathbf N}.$
Because of the specific appearance of
$k_j$ and $k_j^\ast$ in \eqref{5.2} for $1\le j\le \mathbf N,$ the variables 
$\chi_1,\dots,\chi_{\mathbf N}$ appear on those three determinants in a symmetrical way.
Each variable $\chi_j$ appears in the polynomial at most quadratically.
The aforementioned symmetry allows us to determine the coefficients in those polynomials
explicitly in terms of the parameters in the set
$\{k_j,k_j^\ast,\gamma_j,\gamma_j^\ast\}_{j=1}^{\mathbf N}.$
For example, if we determine the coefficient of the term
$\chi_1$ in the polynomial, then we also get the coefficient of each
$\chi_j$ by exploiting the aforementioned symmetry.
By setting $\chi_j=0$ for $2\le j\le \mathbf N,$ from \eqref{5.32} we get
\begin{equation}\label{5.57}
\det[\mathbf M(x)]=\det\begin{bmatrix}
k_1^{2\mathbf N-1}+(zk_1)^{2\mathbf N-1}\gamma_1\,\chi_1 &  \cdots &k_1+zk_1\gamma_1\,\chi_1 &1+\gamma_1\,\chi_1\\
\noalign{\medskip}
(k_1^\ast)^{2\mathbf N-1}+(z^2k_1^\ast)^{2\mathbf N-1}\gamma_1^\ast\,\chi_1& \cdots & k_1^\ast +z^2 k_1^\ast\gamma_1^\ast\chi_1& 1+\gamma_1^\ast\chi_1\\
\noalign{\medskip}
\vdots & \ddots & \vdots & \vdots \\
k_{\mathbf N} ^{2\mathbf N-1}&  \cdots & k_{\mathbf N}& 1\\
\noalign{\medskip}
(k_{\mathbf N}^\ast) ^{2\mathbf N-1}& \cdots & k_{\mathbf N}^\ast & 1
\end{bmatrix},
\end{equation}
which can be written as
\begin{equation}\label{5.58}
\det[\mathbf M(x)]=\alpha_0+\alpha_{11}\,\chi_1+\alpha_{12}\,\chi_1^2,
\end{equation}
where the coefficients $\alpha_0,$ $\alpha_{11},$ and $\alpha_{12}$ 
are
explicitly expressed in terms of the 
parameters in the set
$\{k_j,k_j^\ast,\gamma_j,\gamma_j^\ast\}_{j=1}^{\mathbf N}.$
The coefficient $\alpha_0$ is obtained by letting $\chi_1=0$ in \eqref{5.57}, from which
we have
\begin{equation}\label{5.59}
\alpha_0=\det\begin{bmatrix}
k_1^{2\mathbf N-1} & k_1^{2\mathbf N-2}& \cdots &k_1 &1\\
\noalign{\medskip}
(k_1^\ast)^{2\mathbf N-1}&(k_1^\ast)^{2\mathbf N-2} & \cdots & k_1^\ast & 1\\
\noalign{\medskip}
\vdots & \vdots & \ddots & \vdots & \vdots \\
k_{\mathbf N} ^{2\mathbf N-1}& k_{\mathbf N} ^{2\mathbf N-2}& \cdots & k_{\mathbf N}& 1\\
\noalign{\medskip}
(k_{\mathbf N}^\ast) ^{2\mathbf N-1}& (k_{\mathbf N}^\ast) ^{2\mathbf N-2}& \cdots & k_{\mathbf N}^\ast & 1
\end{bmatrix}.
\end{equation}
The right-hand side in \eqref{5.59} is related to a Vandermonde
determinant, and we write \eqref{5.59} as
\begin{equation}\label{5.60}
\alpha_0=\mathcal V(k_1,k_1^\ast,k_2,k_2^\ast,\cdots,k_{\mathbf N},k_{\mathbf N}^\ast),
\end{equation}
where the Vandermonde coefficient on the right-hand side of \eqref{5.60} is defined as
\begin{equation}\label{5.61}
\mathcal V(a_1,a_2,a_3,a_4,\cdots,a_{2p-1},a_{2p}):=
\prod_{1\le j< l\le 2p} (a_j-a_l).
\end{equation}
We write \eqref{5.60} as
\begin{equation}\label{5.62}
\alpha_0=\mathcal U_{00},
\end{equation}
we recover the coefficient $\alpha_{11}$ appearing in \eqref{5.58} from \eqref{5.57} as
\begin{equation}\label{5.63}
\alpha_{11}=\gamma_1\,\mathcal U_{10}+\gamma_1^\ast\,\mathcal U_{01},
\end{equation}
and we also obtain 
the coefficient $\alpha_{12}$ in \eqref{5.58} as
\begin{equation}\label{5.64}
\alpha_{12}=\gamma_1 \gamma_1^\ast\,\mathcal U_{11},
\end{equation}
where we have defined
\begin{equation}\label{5.65}
\begin{cases}
\mathcal U_{00}:=\mathcal V(k_1,k_1^\ast,k_2,k_2^\ast,\cdots,k_{\mathbf N},k_{\mathbf N}^\ast),\\
\noalign{\medskip}
\mathcal U_{10}:=\mathcal V(zk_1,k_1^\ast,k_2,k_2^\ast,\cdots,k_{\mathbf N},k_{\mathbf N}^\ast),\\
\noalign{\medskip}
\mathcal U_{01}:=\mathcal V(k_1,z^2k_1^\ast,k_2,k_2^\ast,\cdots,k_{\mathbf N},k_{\mathbf N}^\ast),\\
\noalign{\medskip}
\mathcal U_{11}:=\mathcal V(zk_1,z^2k_1^\ast,k_2,k_2^\ast,\cdots,k_{\mathbf N},k_{\mathbf N}^\ast).
\end{cases}
\end{equation}
In a similar manner, by setting 
$\chi_j=0$ for $2\le j\le \mathbf N$ in 
the expression for $\det[\mathbf M_2(x)],$ 
we get
\begin{equation}\label{5.66}
\det[\mathbf M_2(x)]=\beta_0+\beta_{11}\,\chi_1+\beta_{12}\,\chi_1^2.
\end{equation}
During the evaluation of $\det[\mathbf M_2(x)],$ we can factor out certain common terms from the rows
of the matrix $\mathbf M_2(x)$
and then evaluate the determinant of the resulting simplified matrix as a Vandermonde determinant.
Hence, in order to express 
the coefficients $\beta_0,$ $\beta_{11},$ $\beta_{12}$
in \eqref{5.66} 
explicitly in terms of the 
parameters in the set
$\{k_j,k_j^\ast,\gamma_j,\gamma_j^\ast\}_{j=1}^{\mathbf N},$
it is convenient to introduce the modified Vandermonde coefficient
$\tilde{\mathcal V}(a_1,a_2,a_3,a_4,\cdots,a_{2p-1},a_{2p})$ as
\begin{equation*}
\tilde{\mathcal V}(a_1,a_2,a_3,a_4,\cdots,a_{2p-1},a_{2p}):=\left(
\sum_{1\le j< l\le 2p} a_j\,a_l\right)
\mathcal V(a_1,a_2,a_3,a_4,\cdots,a_{2p-1},a_{2p}).
\end{equation*}
With the help of the modified
Vandermonde coefficient, we determine the coefficients $\beta_0,$ $\beta_{11},$ $\beta_{12}$ appearing in \eqref{5.66} as
\begin{equation*}
\beta_0=-\tilde{\mathcal V}(k_1,k_1^\ast,k_2,k_2^\ast,\cdots,k_{\mathbf N},k_{\mathbf N}^\ast),
\end{equation*}
\begin{equation*}
\beta_{11}=-\gamma_1\,\tilde{\mathcal V}(zk_1,k_1^\ast,k_2,k_2^\ast,\cdots,k_{\mathbf N},k_{\mathbf N}^\ast)
-\gamma_1^\ast\,\tilde{\mathcal V}(k_1,z^2k_1^\ast,k_2,k_2^\ast,\cdots,k_{\mathbf N},k_{\mathbf N}^\ast),
\end{equation*}
\begin{equation*}
\beta_{12}=-\gamma_1\,\gamma_1^\ast\,\tilde{\mathcal V}(zk_1,z^2k_1^\ast,k_2,k_2^\ast,\cdots,k_{\mathbf N},k_{\mathbf N}^\ast).
\end{equation*}

As mentioned already, the determinant of
$\mathbf M(x)$
is a polynomial 
of degree $2\mathbf N$
in the
$\mathbf N$ real-valued variables
$\chi_1,\dots,\chi_{\mathbf N}.$
The coefficient of the highest-degree term
$\chi_1^2\chi_2^2\cdots \chi_{\mathbf N}^2$
in $\det[\mathbf M(x)]$ is obtained in a manner similar to
the evaluation of $\alpha_{12}$ given in \eqref{5.64}, and that coefficient
is equal to the quantity given by
\begin{equation}\label{5.71}
\left(\prod_{j=1}^{\mathbf N} \gamma_j\,\gamma_j^\ast\right)\mathcal V(zk_1,z^2 k_1^\ast,zk_2,z^2k_2^\ast,\cdots,zk_{\mathbf N},z^2k_{\mathbf N}^\ast).
\end{equation}
Since $\det[\mathbf M_1(x)]$ is also a polynomial
of degree $2\mathbf N$
 in the
$\mathbf N$ real-valued variables
$\chi_1,\dots,\chi_{\mathbf N},$
as in \eqref{5.58} and \eqref{5.66}, by setting 
$\chi_j=0$ for $2\le j\le \mathbf N$ in 
the expression for $\det[\mathbf M_1(x)],$ 
we obtain
\begin{equation}\label{5.72}
\det[\mathbf M_1(x)]=\epsilon_0+\epsilon_{11}\,\chi_1+\epsilon_{12}\,\chi_1^2.
\end{equation}
With the help of \eqref{5.26}, \eqref{5.54}, and \eqref{5.58}, we evaluate the coefficients
in \eqref{5.72} as
\begin{equation*}
\epsilon_0=\Sigma_{\mathbf N}\,\alpha_0,\quad
\epsilon_{11}=[\Sigma_{\mathbf N} +(z-1)k_1]\alpha_{11},\quad
\epsilon_{12}=[\Sigma_{\mathbf N} +2(z-1)k_1]\alpha_{12},
\end{equation*}
where we recall that
$\Sigma_{\mathbf N},$ $\alpha_{11},$ and $\alpha_{12}$ are the quantities in \eqref{5.4},
\eqref{5.63}, and \eqref{5.64}, respectively.

As already indicated, the symmetrical appearance of the quantities $k_j$ in \eqref{5.2} for $1\le j\le N$ allows us to determine
the coefficients of $\chi_j$ and $\chi_j^2$ in the expressions for
$\det[\mathbf M(x)],$
$\det[\mathbf M_1(x)],$ and $\det[\mathbf M_2(x)].$
For example, with $\chi_j\ne 0$ if we choose the remaining quantities
$\chi_l$ as zero for $l\ne j,$ then we get the analog of \eqref{5.58} as
\begin{equation}\label{5.74}
\det[\mathbf M(x)]=\alpha_0+\alpha_{jj}\,\chi_j+\alpha_{j(j+1)}\,\chi_j^2,
\end{equation}
where $\alpha_{jj}$ and $\alpha_{j(j+1)}$ are the analogs of $\alpha_{11}$ and $\alpha_{12}$ in
\eqref{5.63} and \eqref{5.64}, respectively, and they are expressed as
\begin{equation}\label{5.75}
\alpha_{jj}=\gamma_j\,\mathcal U_{j0}+\gamma_j^\ast\,\mathcal U_{0j},\quad
\alpha_{j(j+1)}=\gamma_j\,\gamma_j^\ast\,\mathcal U_{jj},\qquad 1\le j\le \mathbf N.
\end{equation}
We remark that $\mathcal U_{j0}$ for $2\le j\le \mathbf N$ is defined in a similar way $\mathcal U_{10}$ appearing in the second line of
\eqref{5.65} is defined, i.e. by replacing $k_j$ appearing on the right-hand side of \eqref{5.60} by $zk_j.$
Similarly, $\mathcal U_{0j}$ is defined in an analogous manner $\mathcal U_{01}$ appearing in the third line of
\eqref{5.65} is defined, i.e. by replacing $k_j^\ast$ appearing on the right-hand side of \eqref{5.60} by $z^2k_j^\ast.$
Likewise, $\mathcal U_{jj}$ is defined in a similar way $\mathcal U_{11}$ appearing in the fourth line of
\eqref{5.65} is defined, i.e. by replacing $k_j$ and $k_j^\ast$ appearing on the right-hand side of \eqref{5.60} by $zk_j$ and 
$z^2k_j^\ast,$ respectively.

\section{Soliton solutions to the Sawada--Kotera equation}
\label{section6}

In Section~\ref{section5} we have constructed complex-valued solution pairs
$(Q,P)$ to the integrable system \eqref{2.6}. This is done by solving the inverse scattering
problem for \eqref{2.10} in the reflectionless case by using as input the bound-state poles
of the left transmission coefficient
$T_{\text{\rm{l}}}(k)$
given in \eqref{5.1} and the time-evolved modified
bound-state dependency constants.
By using the input data set
$\{k_j,\gamma_j\}_{j=1}^{\mathbf N},$
with $k_j$ being the quantity in \eqref{5.5} and
$\gamma_j$ being the complex-valued modified bound-state
dependency constant in \eqref{5.21}, we
have constructed the potential pair $(Q,P)$ as in \eqref{5.50} and \eqref{5.51}.
Since we have used the time evolution for the dependency constants as specified in \eqref{4.6} and \eqref{5.24},
the resulting time-evolved potential pair $(Q,P)$ constitutes a solution to the integrable coupled system 
\eqref{2.6}.

In this section, by imposing the appropriate restriction on the solution pair
$(Q,P)$ appearing in \eqref{5.50} and \eqref{5.51},
we construct the $\mathbf N$-soliton solution to the
Sawada--Kotera equation \eqref{2.1}.
It turns out \cite{ACTU2025} that 
relevant
restriction is accomplished by specifying each of the ratios
$s_j/r_j$ for $1\le j\le \mathbf N$ in the input data set
$\{k_j,\gamma_j\}_{j=1}^{\mathbf N}$
appropriately in terms of the elements in the set
$\{k_j\}_{j=1}^{\mathbf N}$ or equivalently in the set
$\{\eta_j\}_{j=1}^{\mathbf N}.$ 
Here, we use $r_j$ and $s_j$ to denote the real and imaginary
parts of $\gamma_j,$ respectively. Thus, we have
\begin{equation}\label{6.1}
\gamma_j=r_j+i s_j,
\qquad 1\le j\le \mathbf N.
\end{equation}
Because of the symmetrical appearance of
the quantities $k_j$ in \eqref{5.2}, for the determination
of $s_j/r_j$ for $1\le j\le \mathbf N,$
it is sufficient to determine the explicit expression only for
$s_1/r_1$ and use the appropriate symmetries to determine the
remaining ratios $s_j/r_j.$ The restrictions on the ratios
$s_j/r_j$ for $1\le j\le \mathbf N$ 
assure that $Q$ is real valued and we have
either $P\equiv 0$ or $P=Q_x.$
The resulting real-valued expression for $Q$ contains
the $2\mathbf N$ real-valued parameters $\eta_j$ and $r_j$
for $1\le j\le\mathbf N,$
where we recall that $\eta_j$ is related to $k_j$ as in \eqref{5.5}.

In the following steps, we show how the real-valued
$\mathbf N$-soliton solution $Q$
to \eqref{2.1} expressed in terms of the parameters in the set $\{\eta_j,r_j\}_{j=1}^{\mathbf N}$
is obtained from the complex-valued solution pair $(Q,P)$ in \eqref{5.50} and \eqref{5.51}
expressed in terms of the parameters in the set $\{\eta_j,r_j,s_j\}_{j=1}^{\mathbf N}.$

\begin{enumerate}

\item[\text{\rm(a)}] 
When $P\equiv 0,$ by integrating both sides of \eqref{5.51} and using \eqref{5.52} and \eqref{5.53}, 
we write \eqref{5.51} in the equivalent form as
\begin{equation}\label{6.2}
\displaystyle\frac{1}{2}\left[A_{2\mathbf N-1}(x)\right]^2 -A'_{2\mathbf N-1}(x)-
A_{2\mathbf N-2}(x)=\Pi_{\mathbf N}-\displaystyle\frac{1}{2}\,\Sigma_{\mathbf N}^2,
\end{equation}
where we recall that we use a prime to denote the $x$-derivative.
Using \eqref{5.42} and \eqref{5.55} in \eqref{6.2}, after some simplifications, we obtain
the equivalent expression given by
\begin{equation}\label{6.3}
\begin{split}
\left(\Pi_{\mathbf N}-\Sigma_{\mathbf N}^2
\right)\det[\mathbf M(x)]+\Sigma_{\mathbf N}
\,\displaystyle\frac{d \det[\mathbf M(x)]}{dx}+
&\displaystyle\frac{d^2\det[\mathbf M(x)]}{dx^2}+
\det[\mathbf M_2(x)]\\
&=\displaystyle\frac{1}{
2\,\det[\mathbf M(x)]}\left(
\displaystyle\frac{d \det[\mathbf M(x)]}{dx}
\right)^2.
\end{split}
\end{equation}

\item[\text{\rm(b)}] 
On the other hand, if $P=Q_x$ then \eqref{5.50} and \eqref{5.51} yield
\begin{equation}\label{6.4}
A_{2\mathbf N-1}(x)\,A'_{2\mathbf N-1}(x)-
A'_{2\mathbf N-2}(x)=0.
\end{equation}
Integrating both sides of \eqref{6.4} and using \eqref{5.50}, \eqref{5.52}, and \eqref{5.53}, 
we write \eqref{6.4} in the equivalent form as
\begin{equation}\label{6.5}
\displaystyle\frac{1}{2}\left[A_{2\mathbf N-1}(x)\right]^2-
A_{2\mathbf N-2}(x)=\Pi_{\mathbf N}-\displaystyle\frac{1}{2}\,\Sigma_{\mathbf N}^2.
\end{equation}
Using \eqref{5.42} and \eqref{5.55} in \eqref{6.5}, after some simplifications, we write
\eqref{6.5} in the equivalent form given by
\begin{equation}\label{6.6}
\begin{split}
\left(\Pi_{\mathbf N}-\Sigma_{\mathbf N}^2
\right)\det[\mathbf M(x)]+\Sigma_{\mathbf N}
\,\displaystyle\frac{d \det[\mathbf M(x)]}{dx}+
&\displaystyle\frac{d^2\det[\mathbf M(x)]}{dx^2}+
\det[\mathbf M_2(x)]\\
&=-\displaystyle\frac{1}{
2\,\det[\mathbf M(x)]}\left(
\displaystyle\frac{d \det[\mathbf M(x)]}{dx}
\right)^2.
\end{split}
\end{equation}

\item[\text{\rm(c)}] From Section~\ref{section5}, we know that each of
$\det[\mathbf M(x)]$
and 
$\det[\mathbf M_2(x)]$
is a polynomial of degree $2\mathbf N$ in the
$\mathbf N$ real-valued variables
$\chi_1,\dots,\chi_{\mathbf N}.$
Consequently, each of
$d\det[\mathbf M(x)]/dx$
and $d^2\det[\mathbf M(x)]/dx^2$
is also a polynomial of degree $2\mathbf N$ in
$\chi_1,\dots,\chi_{\mathbf N}.$ 
Thus, the left-hand sides of \eqref{6.3} and \eqref{6.6} are
some polynomials of degree $2\mathbf N$ in
$\chi_1,\dots,\chi_{\mathbf N}.$
We remark that the left-hand sides of
 \eqref{6.3} and \eqref{6.6} coincide while their right-hand sides differ by a minus sign.
Thus, those right-hand sides 
must also be polynomials of degree $2\mathbf N$ in
$\chi_1,\dots,\chi_{\mathbf N}.$

\item[\text{\rm(d)}] 
In order to satisfy the restrictions that
the right-hand sides of \eqref{6.3} and \eqref{6.6} must each be 
a polynomial of degree $2\mathbf N$ in
$\chi_1,\dots,\chi_{\mathbf N},$ we let
\begin{equation}\label{6.7}
\det[\mathbf M(x)]=\alpha_0\left[\Delta(x)\right]^2,
\end{equation}
where $\alpha_0$ is the constant appearing in \eqref{5.58}--\eqref{5.60}
and $\Delta(x)$ is a polynomial of degree $\mathbf N$ in
$\chi_1,\dots,\chi_{\mathbf N}.$ 
The use of $\alpha_0$ in \eqref{6.7} ensures that 
$\Delta(x)$ is normalized in the sense that
$\Delta(x)=1$ when we let
$\chi_j=0$ for $1\le j\le\mathbf N.$
Comparing \eqref{5.58} with \eqref{6.7}, we see that, if we set $\chi_j=0$ for $2\le j\le\mathbf N,$ then
$\Delta(x)$ must satisfy
\begin{equation}\label{6.8}
[\Delta(x)]^2=1+\displaystyle\frac{\alpha_{11}}{\alpha_0}\,\chi_1+
\displaystyle\frac{\alpha_{12}}{\alpha_0}\,\chi_1^2,
\end{equation}
and hence the coefficients $\alpha_0,$ $\alpha_{11},$ $\alpha_{12}$
appearing in \eqref{5.58} must be related to each other as
\begin{equation}\label{6.9}
\displaystyle\frac{\alpha_{12}}{\alpha_0}=
\left(\displaystyle\frac{\alpha_{11}}{2\,\alpha_0}\right)^2,
\end{equation}
or equivalently as
\begin{equation}\label{6.10}
4\,\alpha_0\,\alpha_{12}=\left(\alpha_{11}\right)^2.
\end{equation}

\item[\text{\rm(e)}] 
From \eqref{5.62}--\eqref{5.65} we observe that the constraint
\eqref{6.10} can be expressed in terms of the
Vandermonde coefficients appearing in \eqref{5.65} as
\begin{equation}\label{6.11}
4\,\mathcal U_{00}\,\mathcal U_{11}\,\gamma_1\,\gamma_1^\ast=
\left(\mathcal U_{10}\,\gamma_1+\mathcal U_{01}\,\gamma_1^\ast 
\right)^2.
\end{equation}
Using \eqref{6.1}
in \eqref{6.11}, we write 
\eqref{6.11} in terms of $r_1$ and $s_1$ as
\begin{equation}\label{6.12}
4\,\mathcal U_{00}\,\mathcal U_{11}\left(r_1^2+s_1^2\right)=
\left[\mathcal U_{10}\,(r_1+i s_1)+\mathcal U_{01}\,(r_1-i s_1)
\right]^2.
\end{equation}
We divide both sides of \eqref{6.12} by $r_1^2$ 
and write the resulting equality in the equivalent form expressed as a quadratic equation in
$s_1/r_1$ as
\begin{equation}\label{6.13}
\begin{split}
\left[4\,\mathcal U_{00}\,\mathcal U_{11}+(
\mathcal U_{10}-\mathcal U_{01})^2
  \right]\left(\displaystyle\frac{s_1}{r_1}\right)^2-&2i
\left(\mathcal U_{10}^2-\mathcal U_{01}^2\right)
\left(\displaystyle\frac{s_1}{r_1}\right)\\
&+
\left[4\,\mathcal U_{00}\,\mathcal U_{11}-(
\mathcal U_{10}+\mathcal U_{01})^2\right]
=0.
\end{split}
\end{equation}
The two solutions to the quadratic equation
\eqref{6.13} are given by
\begin{equation}\label{6.14}
\displaystyle\frac{s_1}{r_1}=
	\displaystyle\frac{i
\,(\mathcal U_{10}^2-\mathcal U_{01}^2)\pm 4\sqrt{\mathcal U_{00}\,\mathcal U_{11} 
\,( \mathcal U_{10}\,\mathcal U_{01}-\mathcal U_{00}\,\mathcal U_{11})} }{ 4\,\mathcal U_{00}\,\mathcal U_{11}+(
\mathcal U_{10}-\mathcal U_{01})^2}.
\end{equation}
From \eqref{5.61} and \eqref{5.66} it follows that
the denominator on the right-hand side of \eqref{6.14} is real and nonzero, the quantity inside the square root 
in the numerator is positive, and $(\mathcal U_{10}^2-\mathcal U_{01}^2)$ is purely
imaginary. Hence, the quadratic equation \eqref{6.13} has two distinct real roots
specified in \eqref{6.14}. In fact, one of them is related to
the constraint regarding the right-hand side of \eqref{6.3}
and the other is related to
the constraint regarding the right-hand side of \eqref{6.6}.

\item[\text{\rm(f)}] By exploiting the symmetries existing in \eqref{5.58} and \eqref{5.74},
with the help of \eqref{5.75}, from \eqref{6.14} we obtain the ratios $s_j/r_j$ as
\begin{equation}\label{6.15}
\displaystyle\frac{s_j}{r_j}=
	\displaystyle\frac{i
\,(\mathcal U_{j0}^2-\mathcal U_{0j}^2)\pm 4\sqrt{\mathcal U_{00}\,\mathcal U_{jj} 
\,( \mathcal U_{j0}\,\mathcal U_{0j}-\mathcal U_{00}\,\mathcal U_{jj})} }{ 4\,\mathcal U_{00}\,\mathcal U_{jj}+(
\mathcal U_{j0}-\mathcal U_{0j})^2},\qquad 1\le j\le \mathbf N.
\end{equation}
As indicated in (e), the two root values appearing on the right-hand side of \eqref{6.15} are real and distinct.

\item[\text{\rm(g)}] We already know that
each of $\mathcal U_{00},$ $\mathcal U_{j0},$ $\mathcal U_{0j},$ $\mathcal U_{jj}$
is determined in terms of the elements in the set
$\{k_j\}_{j=1}^{\mathbf N}$ or equivalently in the set
$\{\eta_j\}_{j=1}^{\mathbf N}$ due to \eqref{5.5}.
From \eqref{6.15} we observe that each ratio $s_j/r_j$ is uniquely expressed in terms of 
the elements in $\{\eta_j\}_{j=1}^{\mathbf N}$
in each of the cases $P\equiv 0$ and $P=Q_x.$
Thus, each of the quantities $(1+i s_j/r_j)$ is a constant uniquely expressed 
 in terms of 
the elements in the set $\{\eta_j\}_{j=1}^{\mathbf N}.$
Let us write \eqref{6.1} as
\begin{equation}\label{6.16}
\gamma_j=r_j\left(1+i\,\displaystyle\frac{s_j}{r_j}\right),
\qquad 1\le j\le \mathbf N.
\end{equation}
Using \eqref{6.15} on the right-hand side of
\eqref{6.16}, we express each $\gamma_j$ as a product of
$r_j$ and a constant
determined by the elements of the set $\{\eta_j\}_{j=1}^{\mathbf N}.$
By using the right-hand side of \eqref{6.16} on the right-hand side of \eqref{5.34}, we express
all entries of the matrix $\mathbf M(x)$ appearing on the right-hand side of \eqref{5.32} in terms of
the elements in $\{\eta_j,r_j\}_{j=1}^{\mathbf N}.$
Using the resulting expression for $\mathbf M(x)$ on the right-hand side of \eqref{5.56}, 
we obtain the solution 
$Q$ to \eqref{2.1} in terms of 
the elements in $\{\eta_j\}_{j=1}^{\mathbf N}.$

\item[\text{\rm(h)}] 
Alternatively, we obtain $Q$
by proceeding as follows.
Using either of the two roots specified in \eqref{6.15}, we uniquely determine
the quantity $\Delta(x)$ in terms of the parameters in the set
$\{k_j,r_j\}_{j=1}^{\mathbf N}$ or equivalently in the set
$\{\eta_j,r_j\}_{j=1}^{\mathbf N}.$ 
Using \eqref{6.7} in \eqref{5.56} we recover the solution $Q$ to
\eqref{2.1} as
\begin{equation}\label{6.17}
Q(x)=6\,\displaystyle\frac{d}{dx}\left(\displaystyle\frac{\Delta'(x)}
{\Delta(x)}\right),
\end{equation}
where we recall that the prime is used for the $x$-derivative and we suppress the $t$-dependence in our notation
$\Delta(x).$

\item[\text{\rm(i)}] 

We recall that $\Delta(x)$ is a real-valued polynomial of degree $\mathbf N$ in the $\mathbf N$ real variables
$\chi_1,\dots,\chi_{\mathbf N}.$ The constant term not containing any of
those $\mathbf N$ variables is equal to $1.$
In each term of $\Delta(x),$ a specific $\chi_j$ either does not appear or appears only to the first power.
From \eqref{5.62}, \eqref{5.63}, \eqref{6.1}, \eqref{6.7}--\eqref{6.10} it follows that 
the coefficient of $\chi_1$ is equal to $\alpha_{11}/(2\alpha_0),$ where
we have
\begin{equation}\label{6.18}
\displaystyle\frac{\alpha_{11}}{2\alpha_0}=\displaystyle\frac{r_1}{2\,\mathcal U_{00}}\left[(\mathcal U_{10}+\mathcal U_{01})+i\,(\mathcal U_{10}-\mathcal U_{01})\,\displaystyle\frac{s_1}{r_1}\right],
\end{equation}
with the understanding that \eqref{6.14} is used for the expression $s_1/r_1$
appearing on the  right-hand side of \eqref{6.18}.
In a similar way, with the help of 
 \eqref{5.74}, \eqref{5.75}, \eqref{6.1}, \eqref{6.7}--\eqref{6.10} it follows that 
the coefficient of $\chi_j$ is equal to $\alpha_{jj}/(2\alpha_0),$ where
we have
\begin{equation}\label{6.19}
\displaystyle\frac{\alpha_{jj}}{2\alpha_0}=\displaystyle\frac{r_j}{2\,\mathcal U_{00}}\left[(\mathcal U_{j0}+\mathcal U_{0j})+i\,(\mathcal U_{j0}-\mathcal U_{0j})\,\displaystyle\frac{s_j}{r_j}\right],\qquad 1\le j\le\mathbf N,
\end{equation}
with the understanding that \eqref{6.15} is used for the expression $s_j/r_j$
appearing on the  right-hand side of \eqref{6.19}. The sign of the free real parameter $r_j$
is chosen so that the right-hand side of \eqref{6.19} is positive.

\item[\text{\rm(j)}] We introduce the quantities $y_j$ as
\begin{equation}\label{6.20}
y_j:=\displaystyle\frac{\alpha_{jj}}{2\alpha_0}\,\chi_j,\qquad 1\le j\le\mathbf N,
\end{equation}
where $\alpha_{jj}/(2\alpha_0)$ is given by the right-hand side of \eqref{6.19} and
we recall that
$\chi_j$ is the quantity in \eqref{5.26}. 
Using \eqref{5.5} in \eqref{5.26} we get
\begin{equation}\label{6.21}
\chi_j=e^{\sqrt{3}\,\eta_j( x-9\eta_j^4 t)}, \qquad 1\le j\le \mathbf N,
\end{equation}
and hence $\chi_j$ remains positive for all real values of $x$ and $t.$
By choosing the sign of each $r_j$ appropriately so that the right-hand side of
\eqref{6.19} is positive, we observe from \eqref{6.20} and \eqref{6.21} that
the quantity $y_j$ defined in \eqref{6.20} remains positive
for $1\le j\le\mathbf N$ for all $x$ and $t.$
We can then write
$\Delta(x)$ as
\begin{equation}\label{6.22}
\begin{split}
\Delta(x)=&1+\sum_{j=1}^{\mathbf N}
y_j+\sum_{1\le j_1<j_2\le \mathbf N} A_{j_1 j_2}\,y_{j_1}\,y_{j_2}\\
&+\sum_{1\le j_1<j_2<j_3\le \mathbf N} A_{j_1 j_2}\,A_{j_1 j_3}\,A_{j_2 j_3}\,
y_{j_1}\,y_{j_2}\,y_{j_3}\\
&+\cdots+
\left[A_{12}\,A_{13}\cdots \,A_{(\mathbf N-1)\mathbf N}\right]
y_1\,y_2\cdots\,y_{\mathbf N},
\end{split}
\end{equation}
where the right-hand side is
a polynomial in $y_1, y_2,\dots, y_{\mathbf N}$
with the last term containing the product $y_1 y_2\cdots y_{\mathbf N}.$
We remark that \eqref{6.22} is the analog of (32) in \cite{P2001}.
With the help of \eqref{5.71} and \eqref{6.7}, 
the double-indexed quantity $A_{jl}$ is determined by the elements
in the set $\{k_j\}_{j=1}^{\mathbf N}$ and is given by
\begin{equation}\label{6.23}
A_{jl}:=\displaystyle\frac{(k_j-k_l)^3(k_j^3+k_l^3)}
{(k_j+k_l)^3(k_j^3-k_l^3)},
\qquad 1\le j<l \le\mathbf N.
\end{equation}
Using \eqref{5.5} in \eqref{6.23}, we can express $A_{jl}$ in terms of the elements of
$\{\eta_j\}_{j=1}^{\mathbf N}$ as
\begin{equation}\label{6.24}
A_{jl}=\displaystyle\frac{(\eta_j-\eta_l)^3(\eta_j^3+\eta_l^3)}
{(\eta_j+\eta_l)^3(\eta_j^3-\eta_l^3)},
\qquad 1\le j<l \le\mathbf N.
\end{equation}
We remark that \eqref{6.24} is the analog of (33) in \cite{P2001}.
Because the $\eta_j$-values are distinct, from \eqref{6.24} it follows that
$A_{jl}$ is positive. Consequently, from \eqref{6.22} we observe that the quantity $\Delta(x)$ remains positive for
all real values of $x$  and $t$ and this assures that the $\mathbf N$-soliton solution $Q$ to \eqref{2.1} appearing in \eqref{6.17} 
and containing the $2\mathbf N$ real-valued parameters
$\eta_j$ and $r_j$ for $1\le j\le \mathbf N$ does not have
any singularities.

\end{enumerate}

\section{Soliton behaviors}
\label{section7}

In Section~\ref{section6} we have presented the construction of the $\mathbf N$-soliton solution to the Sawada--Kotera
equation \eqref{2.1} with the input consisting of the $2\mathbf N$ real parameters in the set $\{\eta_j,r_j\}_{j=1}^{\mathbf N}.$
In this section we elaborate on the method of Section~\ref{section6} when $\mathbf N$ takes the values of $1,$ $2,$ and $3.$
We remark that in  this section we no longer suppress 
the $t$-dependence in the relevant quantities. 

When $\mathbf N=1,$ we use the input data set $\{\eta_1,r_1\}$ to have the $1$-soliton solution $Q(x,t)$ via \eqref{6.17} by
constructing the key quantity $\Delta(x,t)$ appearing in \eqref{6.22}. In this case, \eqref{6.22} reduces to
\begin{equation}\label{7.1}
\Delta(x,t)=1+
y_1,
\end{equation}
where $y_1$ is obtained
from \eqref{6.20} with $\chi_1$ being the quantity we get from \eqref{6.21} when $j=1,$ i.e. we have
 \begin{equation}\label{7.2}
\chi_1=e^{\sqrt{3}\,\eta_1(x-9\eta_1^4 t)}.
\end{equation}
The coefficient $\alpha_{11}/(2\alpha_0)$ in \eqref{6.20} is evaluated with
the help of \eqref{5.65}, \eqref{6.14}, and \eqref{6.18}. By using \eqref{5.65} in \eqref{6.14}, we obtain
\begin{equation}\label{7.3}
\displaystyle\frac{s_1}{r_1}=\begin{cases}
\sqrt{3},\qquad P\equiv 0,\\
\noalign{\medskip}
0,\qquad P=Q_x,
\end{cases}
\end{equation}
where the first line on the right-hand side refers to the case $P\equiv 0$ and the second line
refers to the case $P=Q_x.$
Next, using \eqref{5.65} and \eqref{7.3} in \eqref{6.20}, we obtain the quantity $y_1$ as
\begin{equation}\label{7.4}
y_1=\begin{cases}-2 r_1 \chi_1,\qquad P\equiv 0,\\
\noalign{\medskip}
r_1 \chi_1,\qquad P=Q_x.
\end{cases}
\end{equation}
We remark that either of the two lines on the right-hand side of \eqref{7.4} yields two equivalent values for $y_1$ by restricting the parameter
$r_1$ in the first line to negative values and by
restricting the parameter $r_1$ in the second line to positive values.
Using \eqref{7.1} and \eqref{7.4} in \eqref{6.17},
we obtain the $1$-soliton solution to \eqref{2.1} as
\begin{equation}\label{7.5}
Q(x,t)=\begin{cases} -\displaystyle\frac{36 r_1 \eta_1^2 \chi_1}{\left(1-2 r_1 \chi_1\right)^2},\qquad P\equiv 0,\\
\noalign{\medskip}
\displaystyle\frac{18 r_1 \eta_1^2 \chi_1}{\left(1+r_1 \chi_1\right)^2}   ,\qquad P=Q_x,
\end{cases}
\end{equation}
where it is understood that $r_1<0$ in the first line on the right-hand side and
$r_1>0$ in the second line. We observe that the substitution $r_1\mapsto -r_1/2$ in the first line
yields the second line in \eqref{7.5}. Since $\chi_1$ appearing in \eqref{7.2} is a function of
$(x-9\eta_1^4 t),$ from \eqref{7.5} we see that the $1$-soliton $Q(x,t)$ moves from the left
to the right with the speed $9\eta_1^4.$

\begin{figure}[!ht]
     \centering
       \includegraphics[width=1.45in]{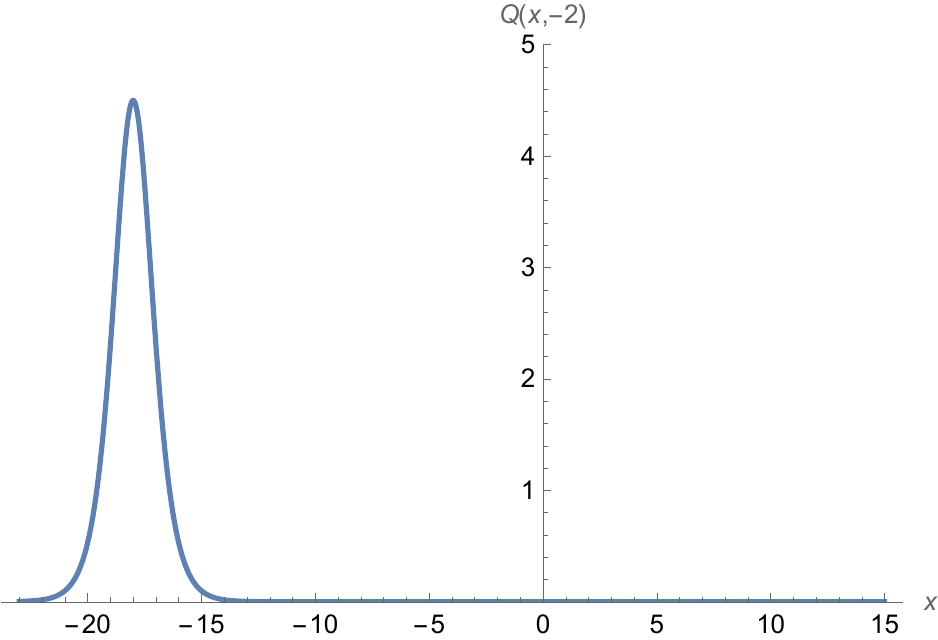}      \hskip .1in
 \includegraphics[width=1.45in]{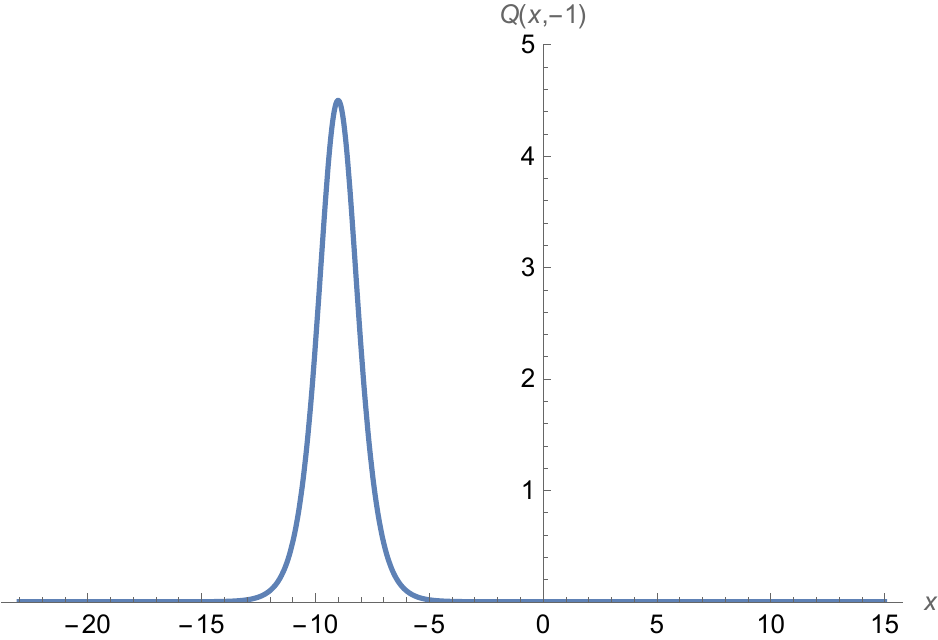} \hskip .1in
\includegraphics[width=1.45in]{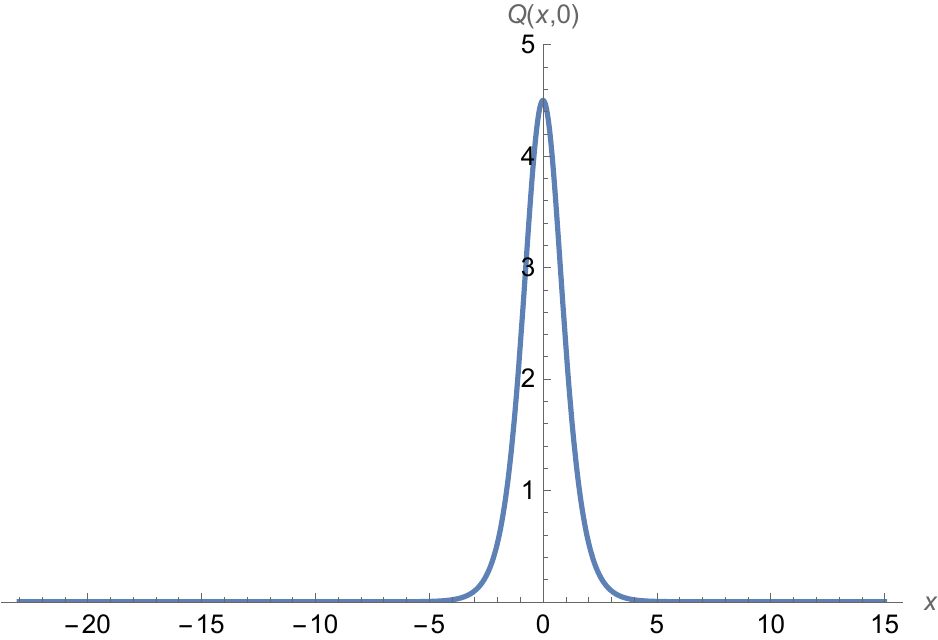} \hskip .1in
\includegraphics[width=1.45in]{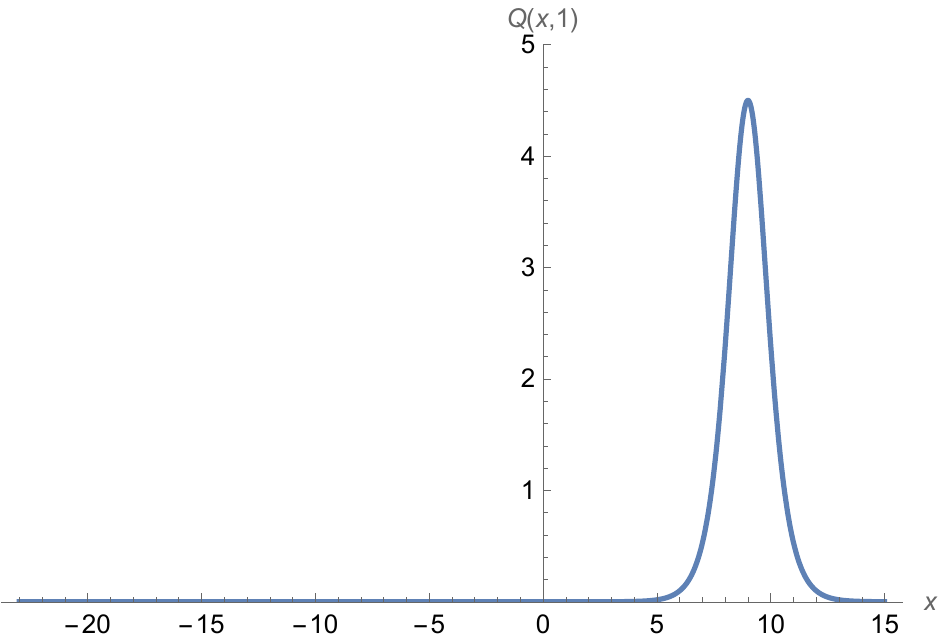} 
\caption{The snapshots for the $1$-soliton solution $Q(x,t)$ to
\eqref{2.1} with the parameter values in \eqref{7.6} at $t=-2,$ $t=-1,$ $t=0,$ and $t=1,$ respectively.}
\label{figure7.1}
\end{figure}

In Figure~\ref{figure7.1} we show the four snapshots at $t=-2,$ $t=-1,$ $t=0,$ and $t=1,$
respectively,
for the  
$1$-soliton solution to \eqref{2.1} constructed with the input parameters given by
\begin{equation}\label{7.6}
(\eta_1,r_1)=\left(1,1\right),
\end{equation}
where we use the second line on the right-hand side of \eqref{7.5} corresponding to $P=Q_x.$ 
From \eqref{7.5} we see that the same soliton solution 
can be constructed by using
the first line of the right-hand side of \eqref{7.5}
with the input parameters
$\eta_1$ and $r_1$ given by
\begin{equation*}
(\eta_1,r_1)=\left(1,-\ds\frac{1}{2}\right),
\end{equation*}
in which case the corresponding $P$ has the value $P(x,t)\equiv 0.$
The $1$-soliton solution $Q(x,t)$ in \eqref{7.5} is a real-valued solitary wave moving from the left to
the right with the constant speed $9\eta_1^4,$ which is equal to $9$ in this particular case.
We remark that the $1$-soliton behavior
for the
Sawada--Kotera equation \eqref{2.1} illustrated in Figure~\ref{figure7.1} is similar
to the behavior 
\cite{AC1991,AS1981,DEGM1982,DJ1989,NMPZ1984}
of the $1$-soliton solution to
the KdV equation in the sense that the soliton moves with a constant speed from the left to the right
without changing its shape.

When $\mathbf N=2,$ we use the input data set $\{\eta_1,\eta_2,r_1,r_2\}$ to
construct the $2$-soliton solution $Q(x,t)$ to
\eqref{2.1} via \eqref{6.17} by forming the quantity $\Delta(x,t)$ appearing in \eqref{6.22}.
In this case,
\eqref{6.22} yields
\begin{equation}\label{7.8}
\Delta(x,t)=1+
y_1+y_2+A_{12}\,y_1\,y_2,
\end{equation}
where $A_{12}$ is obtained from \eqref{6.24} and the
quantities  $y_1$ and $y_2$ are constructed with the help of
\eqref{6.15} and \eqref{6.18}--\eqref{6.21}.
Using \eqref{5.65} in \eqref{6.14}
and exploiting the symmetrical appearance of $k_1$ and $k_2$ in
\eqref{5.2}, we obtain
\begin{equation}\label{7.9}
\displaystyle\frac{s_1}{r_1}=\begin{cases}
\displaystyle\frac{\sqrt{3}\,\eta_2^2}{2\eta_1^2+\eta_2^2},\qquad P\equiv 0,\\
\noalign{\medskip}
-\displaystyle\frac{\sqrt{3}\,\eta_1^2}{\eta_1^2+2\eta_2^2},\qquad P=Q_x,
\end{cases}
\quad
\displaystyle\frac{s_2}{r_2}=\begin{cases}
\displaystyle\frac{\sqrt{3}\,\eta_1^2}{\eta_1^2+2\eta_2^2},\qquad P\equiv 0,\\
\noalign{\medskip}
-\displaystyle\frac{\sqrt{3}\,\eta_2^2}{2\eta_1^2+\eta_2^2},\qquad P=Q_x,
\end{cases}
\end{equation}
where the first lines on the right-hand sides refer to the case $P\equiv 0$ and the second lines
refer to the case $P=Q_x.$
Next, using \eqref{5.65} and \eqref{7.9} in \eqref{6.18}, we obtain the quantity $y_1$
appearing in \eqref{6.20} and \eqref{7.8} as
\begin{equation}\label{7.10}
y_1=\begin{cases} \displaystyle\frac{2 r_1 \left(\eta_1^3+2\eta_1^2 \eta_2+2\eta_1 \eta_2^2+\eta_2^3\right)\chi_1}{\left(\eta_1-\eta_2\right)\left(2\eta_1^2+\eta_2^2\right)},\qquad P\equiv 0,\\
\noalign{\medskip}
-\displaystyle\frac{2 r_1 \left(\eta_1^3+2\eta_1^2 \eta_2+2\eta_1 \eta_2^2+\eta_2^3\right)\chi_1}{\left(\eta_1-\eta_2\right)\left(\eta_1^2+2\eta_2^2\right)},\qquad P=Q_x,
\end{cases}
\end{equation}
with $\chi_1$ being the exponential function given in \eqref{7.2}.
As a result of the ordering $0<\eta_1<\eta_2,$
in the first line on the right-hand side of \eqref{7.10} we use $r_1<0$ and in the second line we use $r_1>0$ so that
the quantity $y_1$ remains positive for all real values of $x$ and $t.$
The value of $y_2$ is obtained with the help of \eqref{7.10} by using the replacement
$(\eta_1,\eta_2,r_1)\mapsto
(\eta_2,\eta_1,r_2).$ Thus, we get
\begin{equation}\label{7.11}
y_2=\begin{cases} -\displaystyle\frac{2 r_2 \left(\eta_1^3+2\eta_1^2 \eta_2+2\eta_1 \eta_2^2+\eta_2^3\right)\chi_2}{\left(\eta_1-\eta_2\right)\left(\eta_1^2+2\eta_2^2\right)},\qquad P\equiv 0,\\
\noalign{\medskip}
\displaystyle\frac{2 r_2 \left(\eta_1^3+2\eta_1^2 \eta_2+2\eta_1 \eta_2^2+\eta_2^3\right)\chi_2}{\left(\eta_1-\eta_2\right)\left(2\eta_1^2+\eta_2^2\right)},\qquad P=Q_x,
\end{cases}
\end{equation}
with $\chi_2$ being the exponential function obtained from \eqref{6.21} when $j=2,$
i.e. we have
 \begin{equation}\label{7.12}
\chi_2=e^{\sqrt{3}\,\eta_2(x-9\eta_2^4 t)}.
\end{equation}
In order to have $y_2$ to remain positive
for all real values of $x$ and $t,$ 
as a result of the ordering
$0<\eta_1<\eta_2,$
we see that
we must use $r_2>0$ in the first line on the right-hand side of \eqref{7.11} and
use $r_2<0$ in the second line.

\begin{figure}[!ht]
     \centering
         \includegraphics[width=1.45in]{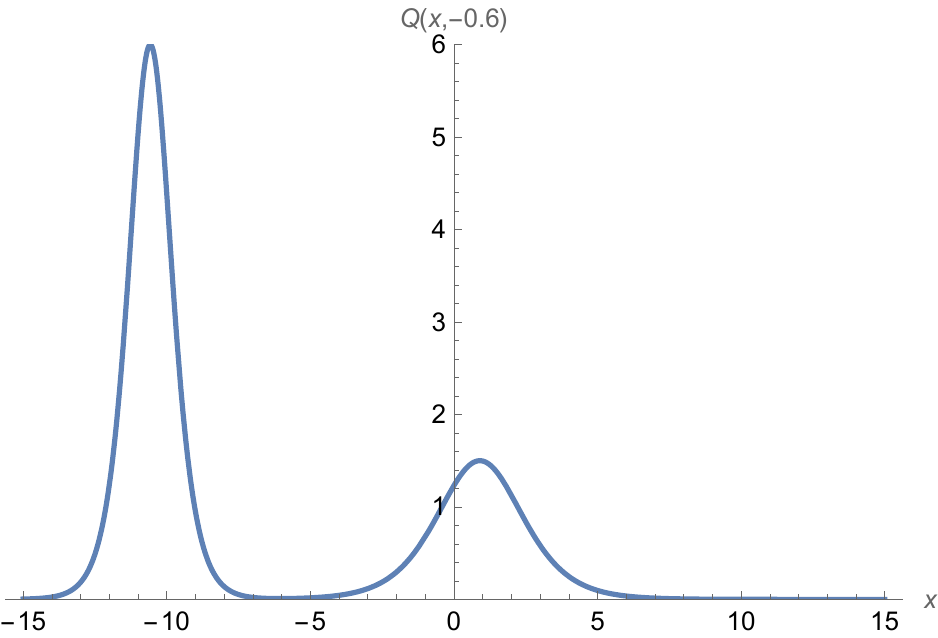}      \hskip .1in
         \includegraphics[width=1.45in]{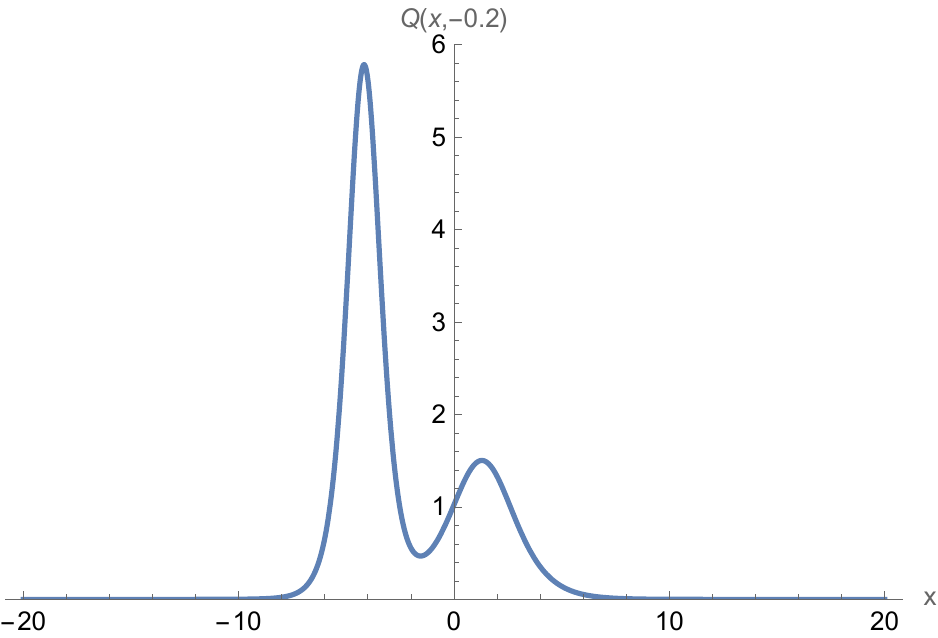} \hskip .1in
           \includegraphics[width=1.45in]{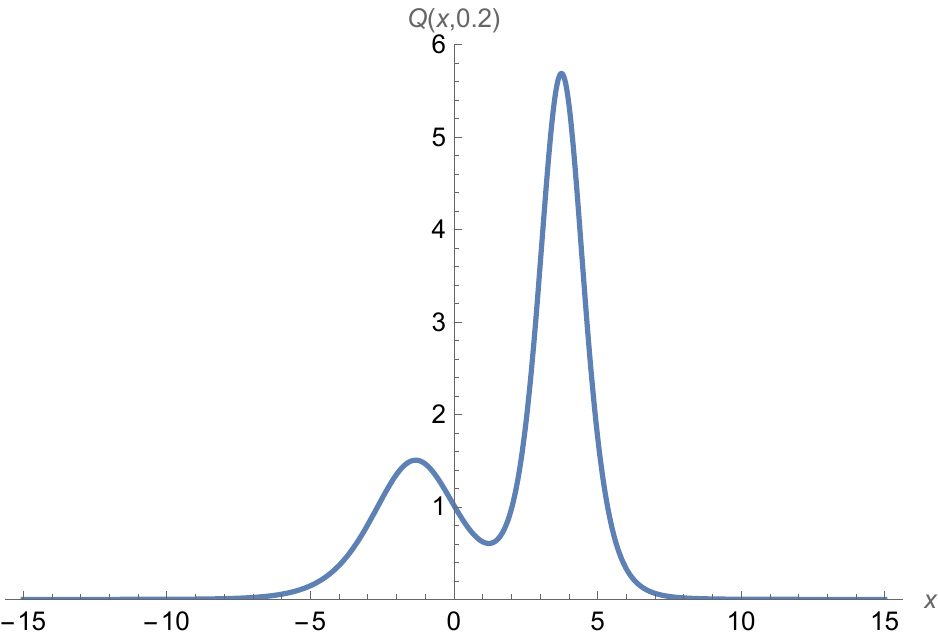} \hskip .1in
            \includegraphics[width=1.45in]{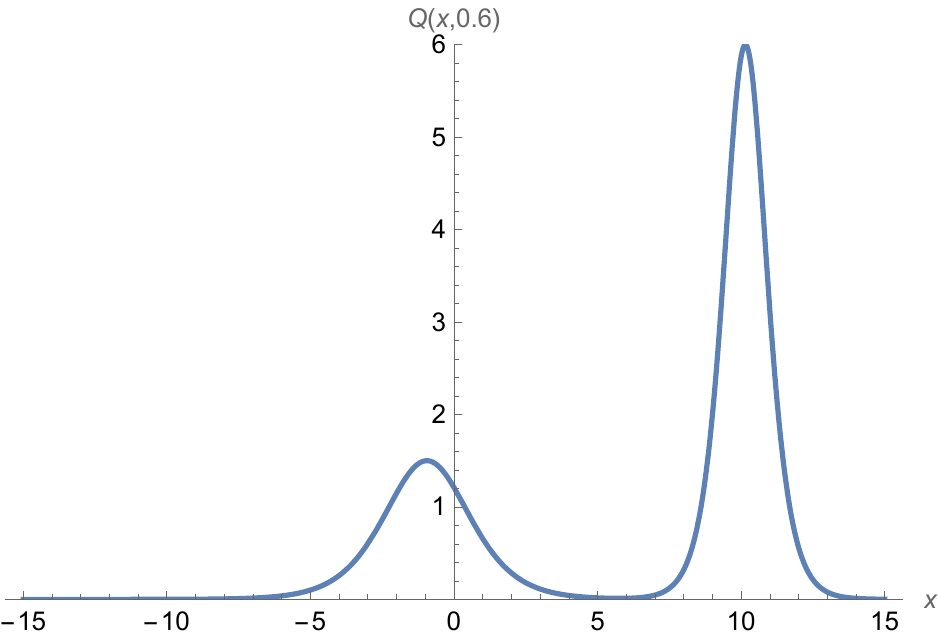} 
\caption{The snapshots for the $2$-soliton solution $Q(x,t)$ to
\eqref{2.1} with the parameter values in \eqref{7.13} at $t=-0.6,$ $t=-0.2,$ $t=0.2,$ and $t=0.6,$ respectively.}
\label{figure7.2}
\end{figure}

In Figure~\ref{figure7.2} we show the behavior of the $2$-soliton solution corresponding to
the input data set 
$\{\eta_1,\eta_2,r_1,r_2\}$ 
with the specific values of the parameters given by
\begin{equation}\label{7.13}
(\eta_1,\eta_2,r_1,r_2)=\left(\displaystyle\frac{1}{\sqrt{3}},\displaystyle\frac{2}{\sqrt{3}},1,-1\right),
\end{equation}
where we use the second lines on the right-hand sides in \eqref{7.9}--\eqref{7.11} corresponding to the case $P=Q_x.$ 
Using the method of Section~\ref{section6}, we explicitly construct
the $2$-soliton solution with the input data set related to \eqref{7.13}. 
Using the values specified in \eqref{7.13} as input to the key quantity  $\Delta(x,t)$ appearing in 
\eqref{7.8}, we obtain
\begin{equation}\label{7.14}
\Delta(x,t)=1+\ds\frac{14}{3}\,\chi_1+
7\,\chi_2+\ds\frac{14}{9}\,\chi_1\chi_2,
\end{equation}
where the quantities $\chi_1$ and $\chi_2$ are constructed with
the help of \eqref{7.2} and \eqref{7.12} as
\begin{equation*}
\chi_1=e^{x-t},\quad \chi_2=e^{2(x-16 t)}.
\end{equation*}
Using \eqref{7.14} in \eqref{6.17}, we obtain the solution $Q(x,t)$ to \eqref{2.1} as
\begin{equation}\label{7.16}
Q(x,t)=\ds\frac{28\,\chi_1+168\,\chi_2+280\,\chi_1\,\chi_2+
\ds\frac{1568}{9}\,\chi_1^2\,\chi_2+
\ds\frac{196}{3}\,\chi_1\,\chi_2^2}{\left(1+\ds\frac{14}{3}\,\chi_1+
7\,\chi_2+\ds\frac{14}{9}\,\chi_1\,\chi_2\right)^2},
\end{equation}
with the understanding that we have $P(x,t)=Q_x(x,t).$
By exploiting the symmetrical appearance of $k_1$ and $k_2$ in \eqref{5.2}, we determine
that we obtain the same $Q(x,t)$ given in \eqref{7.16} with
$P(x,t)\equiv 0$ if we use not the input values in \eqref{7.13} but instead use the input values given by
\begin{equation*}
(\eta_1,\eta_2,r_1,r_2)=\left(\displaystyle\frac{1}{\sqrt{3}},\displaystyle\frac{2}{\sqrt{3}},-\ds\frac{2}{3},
\ds\frac{3}{2}\right).
\end{equation*}

The left plot in 
Figure~\ref{figure7.2} shows the snapshot 
at $t=-0.6$ when the two solitons are apart from each other and not yet interacting with each other. 
The taller soliton has the speed $16$
and is behind the shorter soliton having the speed $1.$
Both solitons move from the left to the right.
The next plot shows the snapshot at $t=-0.2$
soon after the beginning of the nonlinear interactions between the two solitons. The third plot shows the snapshot
at $t=0.2$ when the nonlinear interactions are about to end. Finally, the right plot shows
the snapshot at $t=0.6$ when the two solitons are apart from each other and there are no longer any
interactions. 
By comparing the four snapshots in Figure~\ref{figure7.2}, we observe that, as a result of the
nonlinear interactions, the shorter soliton has been pushed backward
as the taller soliton overtakes that shorter soliton.
During the nonlinear interactions the taller soliton overtakes the shorter soliton.
We also remark that, even though we do not show the snapshot at $t=0$ in Figure~\ref{figure7.2},
the two solitons at $t=0$ in this case overlap in such a way as if there were a single soliton
at that moment.
The $2$-soliton behavior for
the Sawada--Kotera equation is similar to the $2$-soliton behavior
for the KdV equation.

When $\mathbf N=3,$ by using the input data set
$\{\eta_1,\eta_2,\eta_3,r_1,r_2,r_3\},$ 
we construct the $3$-soliton solution to
\eqref{2.1}. As explained in Section~\ref{section6}, we first construct the quantity $\Delta(x,t)$ given in \eqref{6.22} and then
obtain the $3$-soliton solution $Q(x,t)$ via \eqref{6.17}.
When $\mathbf N=3,$ from \eqref{6.22} we get
\begin{equation}\label{7.18}
\Delta(x,t)=1+
y_1+y_2+y_3+
A_{12}\,y_1 \,y_2+ A_{13}\,y_1\, y_3+
A_{23} \,y_2\, y_3+
A_{12}\,A_{13}\,A_{23}
\,y_1\,y_2\,y_3,
\end{equation}
where the coefficients $A_{12},$  $A_{13},$  $A_{23}$
are obtained in terms of $\eta_1,$ $\eta_2,$ $\eta_3$ by using \eqref{6.24}.
We have
\begin{equation*}
A_{12}=\displaystyle\frac{(\eta_1-\eta_2)^3(\eta_1^3+\eta_2^3)}
{(\eta_1+\eta_2)^3(\eta_1^3-\eta_2^3)},\quad
A_{13}=\displaystyle\frac{(\eta_1-\eta_3)^3(\eta_1^3+\eta_3^3)}
{(\eta_1+\eta_3)^3(\eta_1^3-\eta_3^3)},\quad
A_{23}=\displaystyle\frac{(\eta_2-\eta_3)^3(\eta_2^3+\eta_3^3)}
{(\eta_2+\eta_3)^3(\eta_2^3-\eta_3^3)}.
\end{equation*}
We remark that the quantities $A_{13}$ and $A_{23}$ can readily be obtained from
the quantity $A_{12}$ by exploiting the symmetrical appearance of $k_1,$ $k_2,$ and $k_3$ in \eqref{5.2}.
The quantities $y_1,$ $y_2,$ $y_3$ are as in \eqref{6.20}
with $\chi_1,$ $\chi_2,$ $\chi_3$ being the exponential functions appearing in \eqref{6.21}.
Thus, we have
\begin{equation*}
y_1=
\displaystyle\frac{\alpha_{11}}{2\alpha_0}\,\chi_1,\quad
y_2=\displaystyle\frac{\alpha_{22}}{2\alpha_0}\,\chi_2,\quad
y_3=\displaystyle\frac{\alpha_{33}}{2\alpha_0}\,\chi_3,
\end{equation*}
where the three coefficients $\alpha_{jj}/(2\alpha_0)$ for $j=1,2,3$ are obtained
with the help of \eqref{6.19}.
We then construct the $3$-soliton solution $Q(x,t)$ by using \eqref{7.18}
in \eqref{6.17}.

\begin{figure}[!ht]
     \centering
         \includegraphics[width=1.45in]{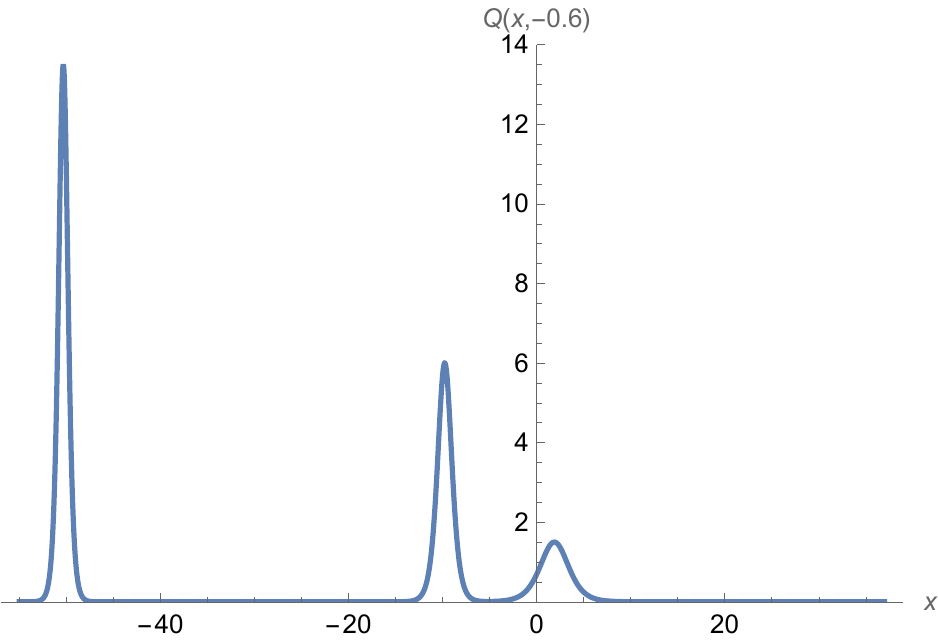}      \hskip .1in
         \includegraphics[width=1.45in]{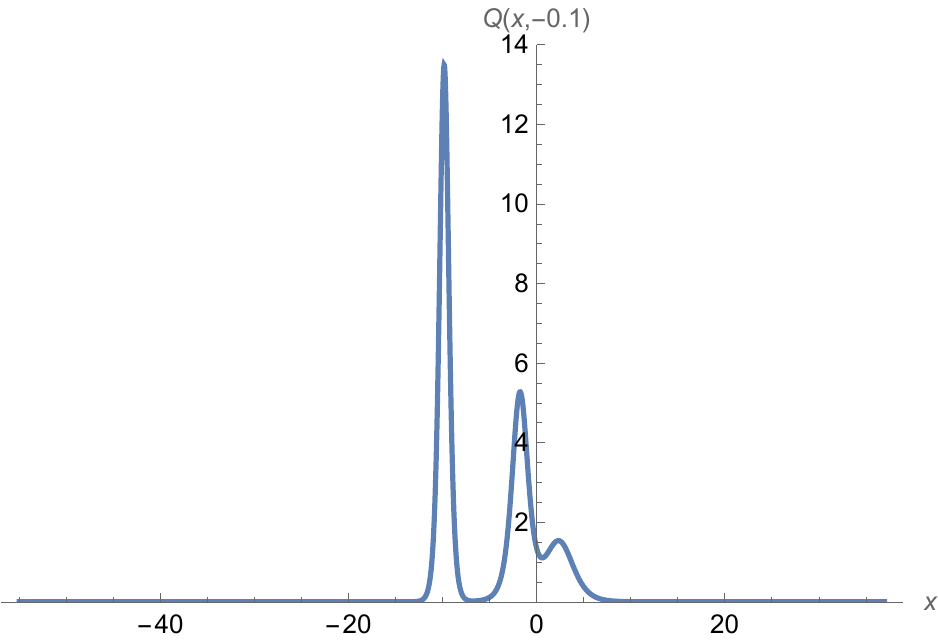} \hskip .1in
           \includegraphics[width=1.45in]{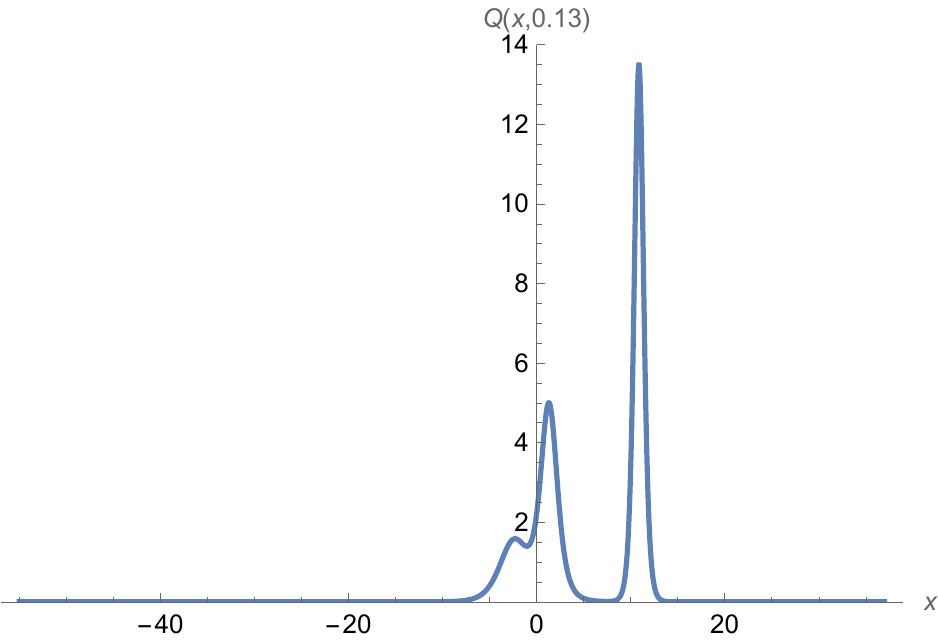} \hskip .1in
            \includegraphics[width=1.45in]{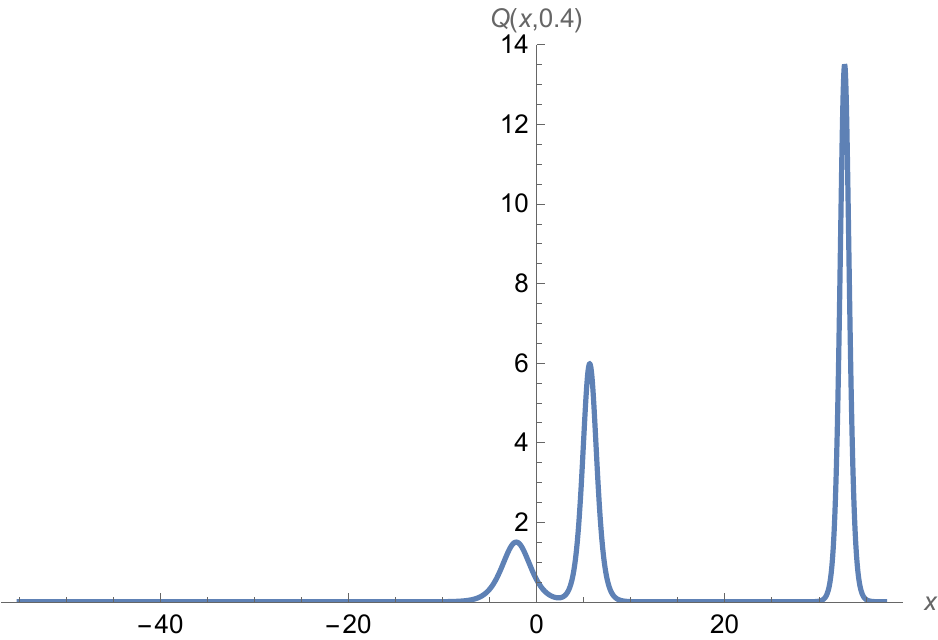} 
\caption{The snapshots for the $3$-soliton solution $Q(x,t)$ to
\eqref{2.1} with the parameter values in \eqref{7.21} at $t=-0.6,$ $t=-0.1,$ $t=0.13,$ and $t=0.4,$ respectively.}
\label{figure7.3}
\end{figure}

In Figure~\ref{figure7.3} we illustrate the behavior of the $3$-soliton solution corresponding to
the input data set $\{\eta_1,\eta_2,\eta_3,r_1,r_2,r_3\}$ with the parameter values given by
\begin{equation}\label{7.21}
(\eta_1,\eta_2,\eta_3,r_1,r_2,r_3)=\left(\displaystyle\frac{1}{\sqrt{3}},\displaystyle\frac{2}{\sqrt{3}},\displaystyle\frac{3}{\sqrt{3}},1,-1,-1\right),
\end{equation}
where we use the case $P=Q_x.$ Via the method of Section~\ref{section6}, we explicitly construct
the $3$-soliton solution with the input data set corresponding to \eqref{7.21}. 
For this, we first use \eqref{7.21} in \eqref{7.18} and obtain the corresponding $\Delta(x,t)$ as
\begin{equation}\label{7.22}
\Delta(x,t)=1+
13\,\chi_1+95\,\chi_2+
\ds\frac{1235}{7}\,\chi_3+
\ds\frac{1235}{21}\,\chi_1\,\chi_2+
\ds\frac{1235}{4}\,\chi_1\,\chi_3+
247\,\chi_2\,\chi_3+
\ds\frac{247}{12}\,\chi_1\,\chi_2\,\chi_3,
\end{equation}
where the quantities $\chi_1,$ $\chi_2,$ $\chi_3$ are obtained with the help of
\eqref{6.21} and \eqref{7.21} as
\begin{equation*}
\chi_1=e^{x-t},
\quad \chi_2=e^{2(x-16 t)},
\quad \chi_3=e^{3(x-81 t)}.
\end{equation*}
The $3$-soliton solution $Q(x,t)$ to \eqref{2.1} is then constructed by using \eqref{7.22} in \eqref{6.17}.

The left plot in 
Figure~\ref{figure7.3} shows the snapshot at $t=-0.6$ when the three solitons are apart from each other and not yet interacting with each other. The next plot shows the snapshot at $t=-0.1$ near
the beginning of the nonlinear interactions among the three solitons. The third plot shows the snapshot
at $t=0.13$ near the end of the nonlinear interactions. Finally, the right plot shows
the snapshot at $t=0.4$ when the three solitons are apart from each other and they no longer interact.
We also observe the following. All three solitons travel from the left to the right. Initially, before the nonlinear interactions start,
they are aligned in such a way that the tallest soliton is behind the other two solitons, the shortest soliton is ahead of the other two, and the
middle-height soliton is between the other two. Prior to any nonlinear interactions,
the tallest soliton has the speed 81, the middle-height soliton 
has the speed $16,$ and the shortest soliton has the speed $1.$
During the nonlinear interactions, the tallest soliton overtakes the other two solitons and the middle-height soliton overtakes the shortest soliton. 
We further observe that the shortest soliton has been pushed backward as a result of the nonlinear interactions.
In Figure~\ref{figure7.3}
 we do not show the snapshot at $t=0,$ but we remark that
the three solitons at $t=0$ in this case overlap in such a way as if there were a single soliton
at that moment.
The $3$-soliton behavior for the Sawada--Kotera equation is similar to
the $3$-soliton behavior for the KdV equation.

We recall that the input data set with the parameter values in \eqref{7.21} yields the quantity
$\Delta(x,t)$ in \eqref{7.22} and the corresponding $3$-soliton solution $Q(x,t)$ 
to \eqref{2.1} via \eqref{6.17}, where the associated quantity $P(x,t)$ is related to
$Q(x,t)$ as
$P(x,t)=Q_x(x,t).$
By exploiting the symmetrical appearance of $k_1,$ $k_2,$ and $k_3$ in \eqref{5.2}, we
construct the same $\Delta(x,t)$ in \eqref{7.22} and hence
the same $Q(x,t)$ with the associated quantity $P(x,t)$ satisfying
$P(x,t)\equiv 0$ if we use not the input values in \eqref{7.21} but instead use the input values given by
\begin{equation*}
(\eta_1,\eta_2,\eta_3,r_1,r_2,r_3)=\left(\displaystyle\frac{1}{\sqrt{3}},\displaystyle\frac{2}{\sqrt{3}},\displaystyle\frac{3}{\sqrt{3}},-\ds\frac{3}{4},\ds\frac{5}{2},-\ds\frac{25}{4}\right).
\end{equation*}

\end{document}